\documentclass{ptephy_v1_arxiv}[dvipdfmx]
\usepackage[utf8]{inputenc}
\usepackage{type1cm}
\usepackage{lmodern}

\title{Performance evaluation of electron multiplier tubes as a high-intensity muon beam monitor of accelerator neutrino experiments}
\author[1,*]{Takashi Honjo}
\author[2]{Yosuke Ashida}
\author[3,4]{Oderich F. Auersperg-Castell}
\author[5]{Megan Friend}
\author[3]{Ian Heitkamp}
\author[3]{Atsuko K. Ichikawa}
\author[6]{Masaki Ishitsuka}
\author[6]{Nao Izumi}
\author[3]{Sohei Kasama\thanks{currently at Collaborative Laboratories for Advanced Decommissioning Science(CLADS), JAEA, Ibaraki, Japan}}
\author[7]{Shigeru Kashiwagi}
\author[1]{Yuma Kawamura}
\author[2]{Tatsuya Kikawa}
\author[1]{Takuya Kobata}
\author[5]{Tsunayuki Matsubara}
\author[7]{Manabu Miyabe}
\author[3]{Kiseki D. Nakamura}
\author[6]{Hina Nakamura}
\author[6]{Yukine Sato}
\author[5]{Ken Sakashita}
\author[1,8]{Yoshihiro Seiya}
\author[3]{Kouichi Takifuji}
\author[7]{Atsushi Tokiyasu}
\author[1]{Tatsuya Yamamoto}
\author[1,8]{Kazuhiro Yamamoto}
\author[2]{Kenji Yasutome\thanks{currently at RIKEN SPring-8 Center (RSC), Kouto, Sayo, Hyogo 679-5148, Japan}}

\affil[1]{Department of Physics, Osaka Metropolitan University, 3-3-138 Sugimoto, Sumiyoshi-ku, Osaka, 558-8585, Japan\email{honjo.ocu.hep@gmail.com}}
\affil[2]{Department of Physics, Kyoto University, Oiwake-cho, Sakyo-ku, Kyoto, 606-8501, Japan}
\affil[3]{Department of Physics, Faculty of Science, Tohoku University, 6-3 Aramakiaza Aoba, Aoba-ku, Sendai, Miyagi, 980-8578, Japan}
\affil[4]{Faculty of Physics, University of Vienna, Boltzmanngasse 5, 1090, Vienna, Austria}
\affil[5]{High Energy Accelerator Research Organization (KEK), Oho, Tsukuba-shi, Ibaraki, 305-0801, Japan}
\affil[6]{Department of Physics, Faculty of Science and Technology, Tokyo University of Science, Yamazaki 4671, Noda, Chiba, 278-8510, Japan}
\affil[7]{Research Center for Accelerator and Radioisotope Science, Tohoku University, 1-2-1 Mikamine, Taihaku-ku, Sendai, 982-0826, Japan}
\affil[8]{Nambu Yoichiro Institute of Theoretical and Experimental Physics (NITEP), 3-3-138 Sugimoto, Sumiyoshi-ku, Osaka, 558-8585, Japan}

\begin{document}
\begin{abstract}
Upgrade work towards increasing the beam intensity of the neutrino beamline at J-PARC is underway.
Monitoring tertiary muon beams is essential for stable operation of the beamline.
Accordingly, we plan to replace the present muon monitor sensors with electron multiplier tubes (EMTs).
We investigated the radiation tolerance and linearity response of EMTs using a 90 MeV electron beam.
An EMTs was irradiated with electrons up to 470 nC.
EMTs show higher radiation tolerance than the Si sensors which are presently used as one of the muon monitor detectors for the T2K long-baseline neutrino experiment at J-PARC.
The integrated charge yield decrease is found to be less than 8\% after a beam irradiation equivalent to 132 days of operation at the future J-PARC beam power of 1.3 MW.
The EMTs show linearity better than ±5\% up to the future beam intensity.
The observed yield decrease is likely due to dynode deterioration based on the detailed investigation.
The studies described here confirm that EMTs can be used as a high-intensity muon beam monitor.
From the reported results, we are proceeding with the installation in the J-PARC neutrino beamline.
\end{abstract}


\maketitle

\section{Introduction}\label{sec:section_Introduction}
In the long-baseline neutrino experiment, it is essential to increase the number of observed neutrino statistics.
At the Japan Proton Accelerator Research Complex (J-PARC)~\cite{J-PARCUpgrade2019}, the Tokai-to-Kamioka (T2K) experiment~\cite{T2K2020} is running and Hyper-Kamiokande~\cite{HK2018} will run to search for CP violation in the lepton sector, and an upgrade to increase the intensity of proton beams for producing neutrino beams is underway.
Neutrinos are produced along with muons from charged pions emitted by protons hitting a graphite target.
Three magnetic horns are installed in the beamline to focus parallel charged pions to the direction of the T2K far detector, Super-Kamiokande~\cite{ICHIKAWA201227}.
Monitoring beam intensity and direction is crucial for maintaining safety at the facility as well as the stable data taking and control of systematic uncertainties in the physics analysis.
In T2K, for the purpose of monitoring the beam intensity and direction, a muon beam monitor (MUMON), composed of two arrays of silicon pin-photodiodes (Si) and ionization chambers (IC), has been used since the beginning of the operation~\cite{MUMON2010,MUMON2015}. 
Each sensor array has $\rm49\, (= 7 \times 7)$ sensors at 25 cm intervals and covers an area of $\rm150 \times 150\, cm^2$ with respect to the beam axis. 
MUMON measures the direction of the muon beam and hence the neutrino beam as well with a precision better than $\rm0.28\, mrad$. 
The resolution is better than $\rm3.0\, mm$, which is equivalent to $\rm0.025\, mrad$.
It also monitors the stability of the beam intensity with a resolution better than 3\%.
To preserve the required precision, it is estimated from a Monte Carlo simulation study that the tolerance for non-linearity is 5\%~\cite{J-PARCUpgrade2019}.
As the proton beam power increases from $\rm510\, kW$ as of 2020 to $\rm1.3\, MW$ in 2028~\cite{PTEP_Accelerator_design}, MUMON needs to be able to work with such high-intensity beams.
This intensity increase is achieved by an increase in the repetition cycle and the number of protons contained per pulse.
Eight bunches of protons, called a spill, are accelerated in the J-PARC Main Ring, and will be extracted every $\rm1.16\, s$ to be transported to the graphite target.
The high-intensity neutrino and muon beams are made by increasing the intensity of the proton beam and the electromagnetic horn current.
The muon flux at the MUMON is estimated from a Monte Carlo simulation to be
$1.09\times10^{5}\, \rm muons/cm^2/(10^{12}\, protons\ on\ target)$ at a +250\, kA horn current.
The muon flux is estimated to be
$8.1\times10^6\,\rm muons/cm^2/bunch$ and $5.6\times10^7\,\rm muons/cm^2/sec$ at MUMON array center with a $\rm1.3\, MW$ proton beam power and a $\rm+320\, kA$ horn current. 
In this study, the focus is both the response to the beam in one bunch and the tolerance to the integrated beam.
Table~\ref{tab:T2Kbeamsetup} summarizes the beam upgrade planned at J-PARC.

It is previously reported that there are concerns about Si and IC regarding their performance at the higher-intensity beam operation; Si yield was found to degrade after irradiation and IC showed a non-linear behavior in relation to beam power~\cite{Ashida2018}.
Diamond sensors were investigated as a candidate of the replacement, but they do not show sufficient radiation tolerance~\cite{Yasutome-san-proceedings}.
As an alternative MUMON sensor candidate in the future, the first prototypes of electron multiplier tubes (EMTs) were investigated using the T2K muon beam~\cite{Ashida2018}. 
This study revealed that the EMT time response is faster than Si and IC, and that they keep signal linearity up to a $\rm460\, kW$ beam power that was the highest power achievable at that time.
The signal yield was stable within ±1\% while an yield drift was observed immediately after the beam irradiation, which is referred to as initial instability in the present paper as one of the main study topics.
These observation indicates the potentiality of EMTs as a new MUMON candidate.
Linear response and radiation tolerance are required to use EMTs at the same position as Si in J-PARC for the next-generation MUMON.
We therefore tested EMTs in detail using an electron beam at Research Center for Accelerator and Radioisotope Science (RARIS) of Tohoku University.
In the beam irradiation test, we investigated the linearity, radiation tolerance, and initial instability of EMTs. 
In addition, we investigated the causes of the observed degradation due to radiation and potential causes for the initial instability.

\begin{table}[tb]
\caption{\label{tab:T2Kbeamsetup}Past and planned parameters of the J-PARC neutrino beamline.
The horn current factor is the muon flux at the MUMON center normalized to the one corresponding to the horn current of $\rm+250\, kA$.
The horn current factor in the future for the sum of all 49 channels is 1.5. In case of the reverse horn current setting, the muon flux decreases as shown in Ref.~\cite{Ashida2018}.}
\centering
\begin{tabular}{lcc}
\hline
\begin{tabular}{l} \end{tabular} Year&2020&2028\\\hline
\begin{tabular}{l}Proton beam power (MW)\end{tabular}&0.51&1.3\\ 
\begin{tabular}{l}Repetition cycle (sec/spill)\end{tabular}&2.48&1.16\\ 
\begin{tabular}{l}Number of protons per spill~\cite{PTEP_Accelerator_design}\end{tabular}& $\text{2.6×10}^\text{14}$ & $\text{3.3×10}^\text{14}$\\ 
\begin{tabular}{l}Horn current (kA)\end{tabular}&+250&+320\\ 
\begin{tabular}{l}Horn current factor at the MUMON center\end{tabular}&1&1.8\\ 
\begin{tabular}{l}Muon flux at the MUMON center\\ (muons$\text{/cm}^\text{2}$/bunch)\end{tabular}&3.5×$\text{10}^\text{6}$&8.1×$\text{10}^\text{6}$\\ 
\begin{tabular}{l}Muon flux at the MUMON center\\(muons$\text{/cm}^\text{2}$/sec)\end{tabular}&1.1×$\text{10}^\text{7}$&5.6×$\text{10}^\text{7}$\\
\hline
\end{tabular}
\end{table}



The rest of this paper is structured as follows. 
After describing our EMT design in detail in Section~\ref{sec:section_EMT}, we show electron beam irradiation tests to evaluate the performance of EMTs in Section ~\ref{sec:section_beam_test}.
In Section~\ref{sec:section_Result}, the results of the beam irradiation tests are presented.
In Section~\ref{sec:section_discussion}, we discuss temperature dependence and causes of radiation degradation, followed by summary in Section~\ref{sec:section_summary}.

\section{Electron Multiplier Tubes}\label{sec:section_EMT}
A photomultiplier tubes (PMT) generate signal when photons enter the glass window with a photocathode. 
They can also generate signal when charged particles penetrate their electrode and secondary electrons are emitted.
The performance of PMTs are degraded by radiation mainly due to degradation of the transmittance of the glass window.
Hence, the radiation tolerance as a charged particle detector using parts besides the glass window is expected to be high.
PMTs were then tested as a candidate sensor, but they were not satisfactory for the use as MUMON since the yields were continuously decreased and they showed a poor linearity~\cite{J-PARCUpgrade2019}.
Because the observed yield decrease seemed to be caused at the photocathode, we decided to replace it with aluminum to achieve higher radiation tolerance. 
HAMAMATSU R9880U series PMTs were modified with their photocathode deposited with aluminum to produce EMTs.
The sensitive area of the aluminized cathode is $50\,\rm mm^2$ and that of the dynode is about $77\,\rm mm^2$.
The secondary electrons produced by charged particles passing through the cathode or dynode are amplified in dynodes. 
The shape of the EMT dynode is a metal channel type with 10 stages.
\par
\begin{figure}[bt]
  \begin{minipage}[b]{0.45\linewidth}
    \centering
    \includegraphics[width=0.7\linewidth]{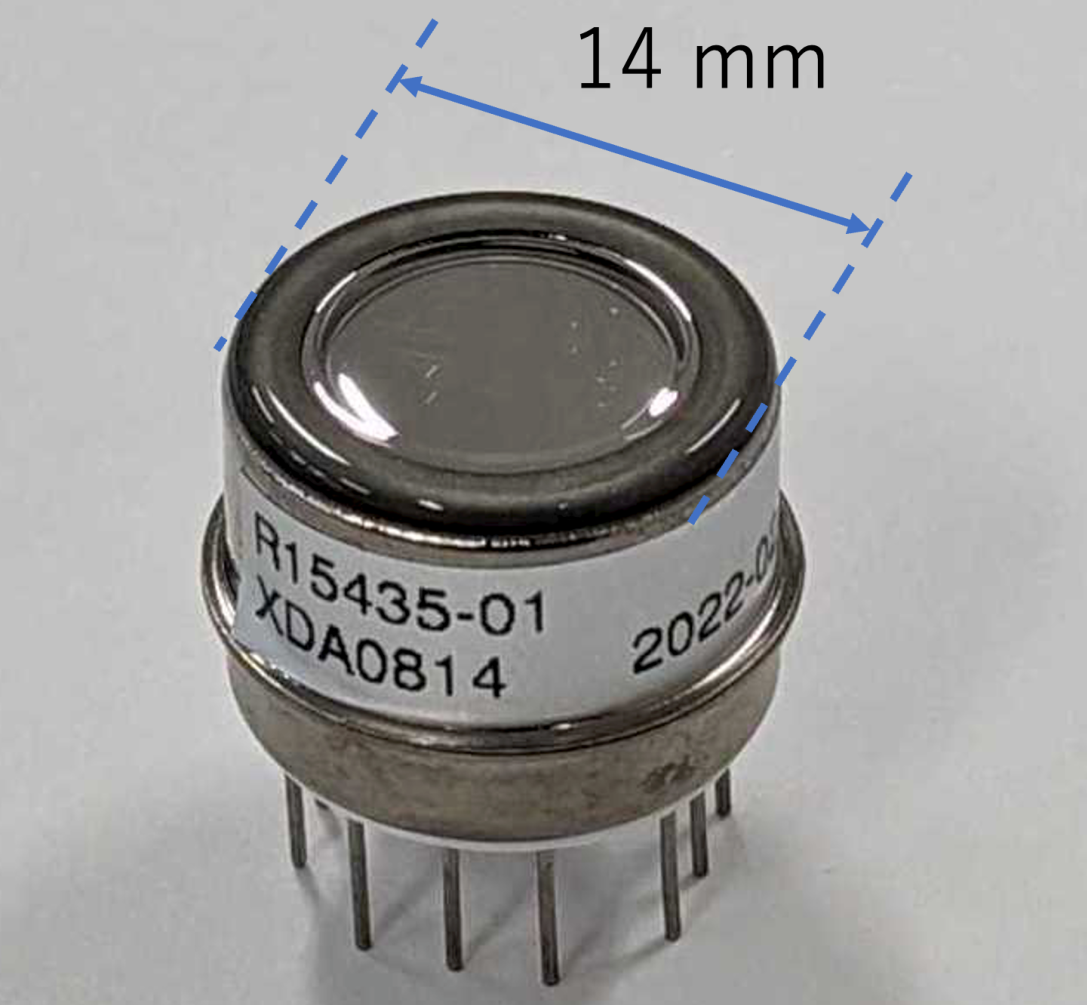}
  \end{minipage}
  \begin{minipage}[b]{0.45\linewidth}
    \centering
    \includegraphics[width=0.755\linewidth]{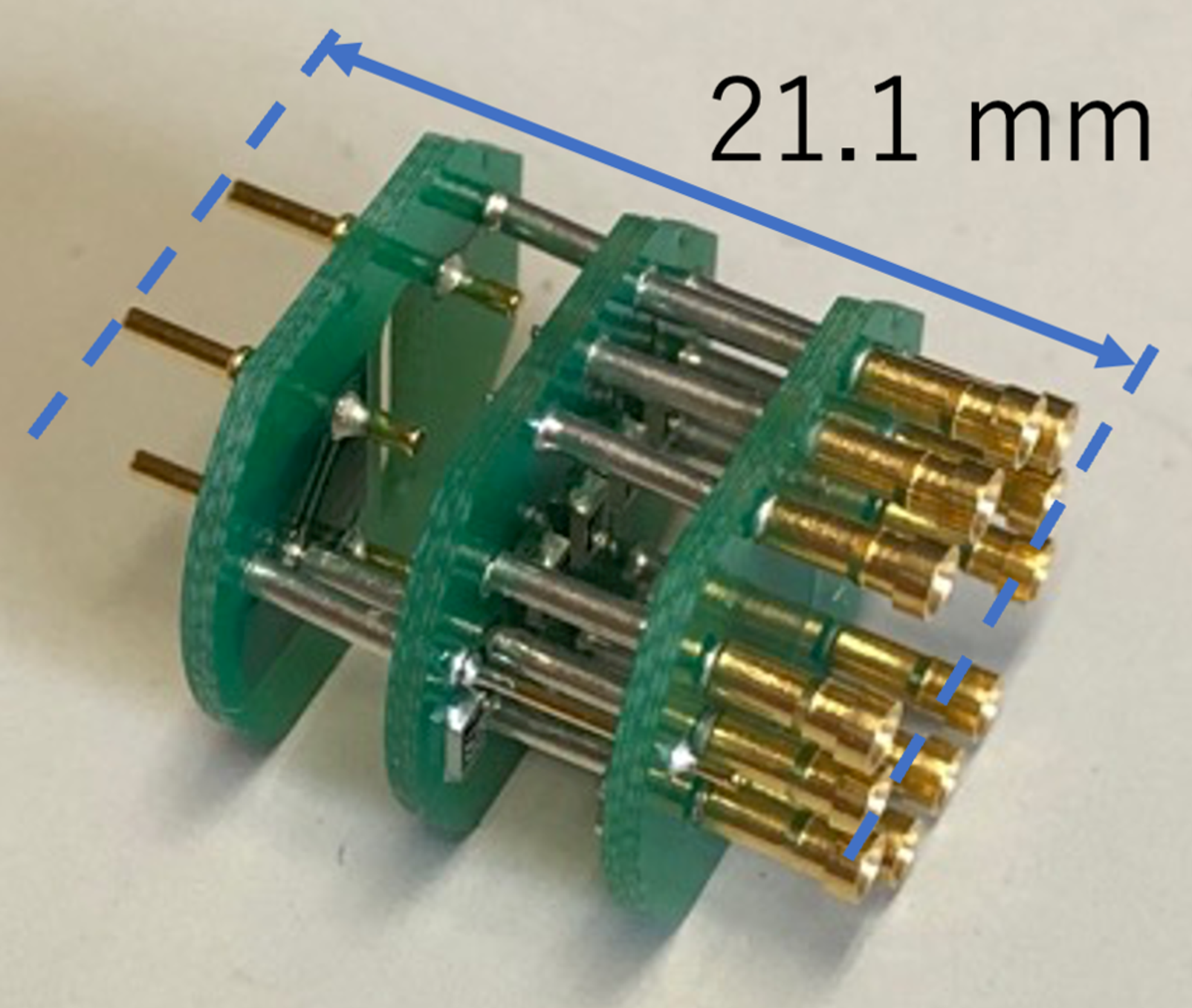}
  \end{minipage}
  \caption{Photograph of an EMT (left) and a bleeder circuit (right).\label{fig:EMTdivider}}
\end{figure}
\begin{figure}[bt]
  \centering
  \includegraphics[width=0.7\linewidth]{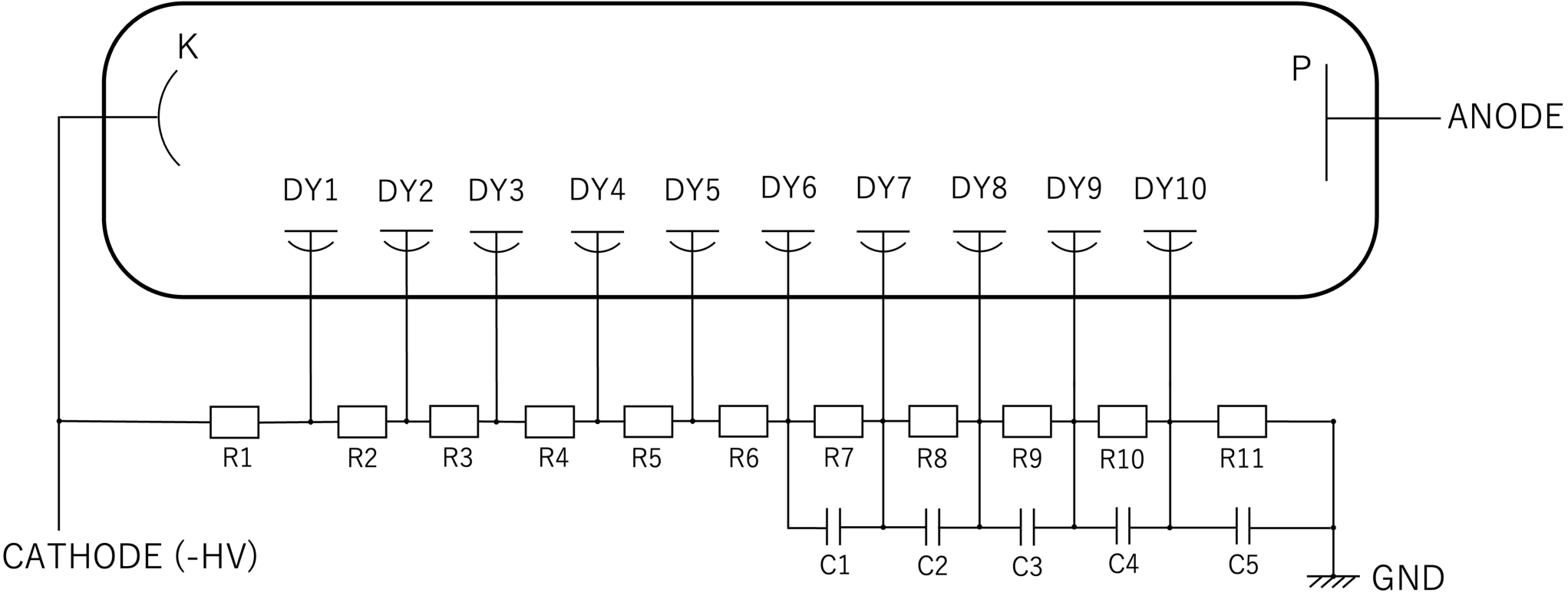}
  \caption{Structure of the bleeder circuit. ``K'', ``P'', ``HV'', ``DY'', ``R'', ``C'' and ``GND'' represent the cathode, anode, high voltage, dynode, resistance, capacitor, and ground, respectively.\label{fig:div_cir}} 
\end{figure}

The original bleeder circuit used for PMTs has an insulating cover made of polybutylene terephthalate (PBT) or polyacetate (POM), but we decided not to use the cover as shown in Figure~\ref{fig:EMTdivider} to avoid potential degradation due to radiation.
The bleeder circuit connected to EMTs consists of eleven resistors $R_1 - R_{11}$ and five capacitors $C_1 - C_5$, as shown in Figure~\ref{fig:div_cir}. 
Each resistance and capacitance values are optimized to maintain a good linearity up to large signal output and are listed in Table~\ref{tab:resistance}.

\begin{table}[bt]
    \caption{The resistance and capacitance values of the bleeder circuit shown in Figure~\ref{fig:div_cir}.}
    \label{tab:resistance}
    \centering
    \begin{tabular}{cc|cc}
        \hline
        &Resistance ($\rm k\Omega$)& &Capacitance ($\mu$F)\\ 
        \hline
        $R_1$ &200 & - & - \\
        $R_2$ &200 & - & -\\
        $R_3$ &150 & - & - \\
        $R_4$ &150 & - & - \\
        $R_5$ &150 & - & - \\
        $R_6$ &150 & - & - \\
        $R_7$ &150 &$C_1$ &0.01 \\
        $R_8$ &150 &$C_2$ &0.01 \\
        $R_9$ &200 &$C_3$ &0.01 \\
        $R_{10}$ &510 &$C_4$ &0.33 \\
        $R_{11}$ & 75 &$C_5$ &0.33 \\
        \hline
    \end{tabular}
\end{table}

\section{Electron Beam Irradiation Test at RARIS}\label{sec:section_beam_test}
\subsection{Experiment}\label{sec:subsection_Setup}
Investigation items in the electron beam irradiation test include linearity, radiation tolerance, and initial instability of EMTs.
The experiment was conducted four times between 2019 and 2022.
In the first beam test in 2019, we mainly tested EMT's linearity. In the second beam test in 2020, we irradiated EMTs as well as Si sensors to compare their radiation tolerances. The third and fourth beam tests were performed in 2021 and 2022, respectively, to investigate specific features of EMTs and their individual difference in quality. In particular, we investigated temperature dependence of the EMT output.
\par
\begin{figure}[bt]
  \centering
  \begin{minipage}[b]{0.4\linewidth}
    \includegraphics[width=0.75\linewidth]{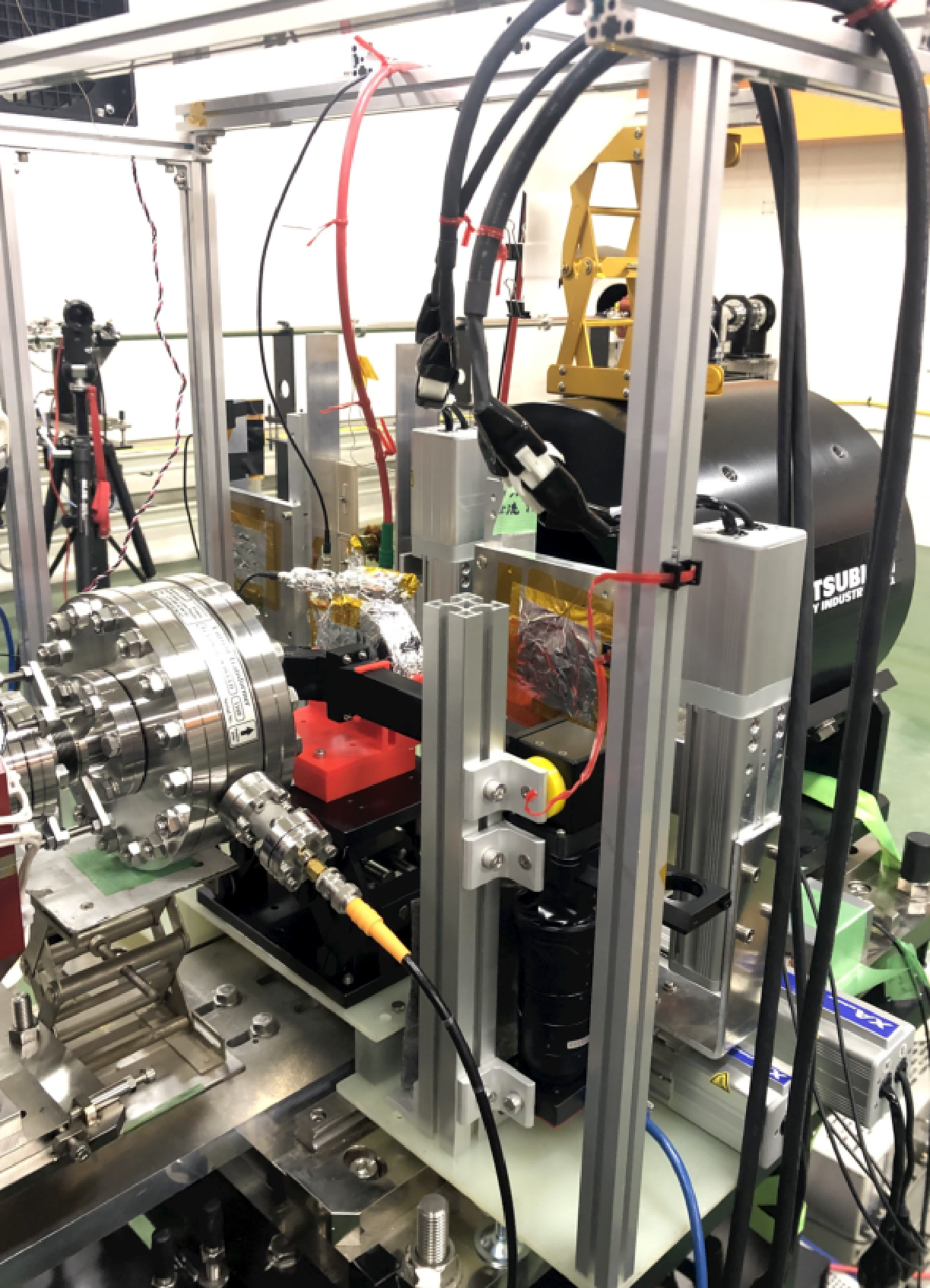}
  \end{minipage}
  \begin{minipage}[b]{0.5\linewidth}
    \includegraphics[width=1.0\linewidth]{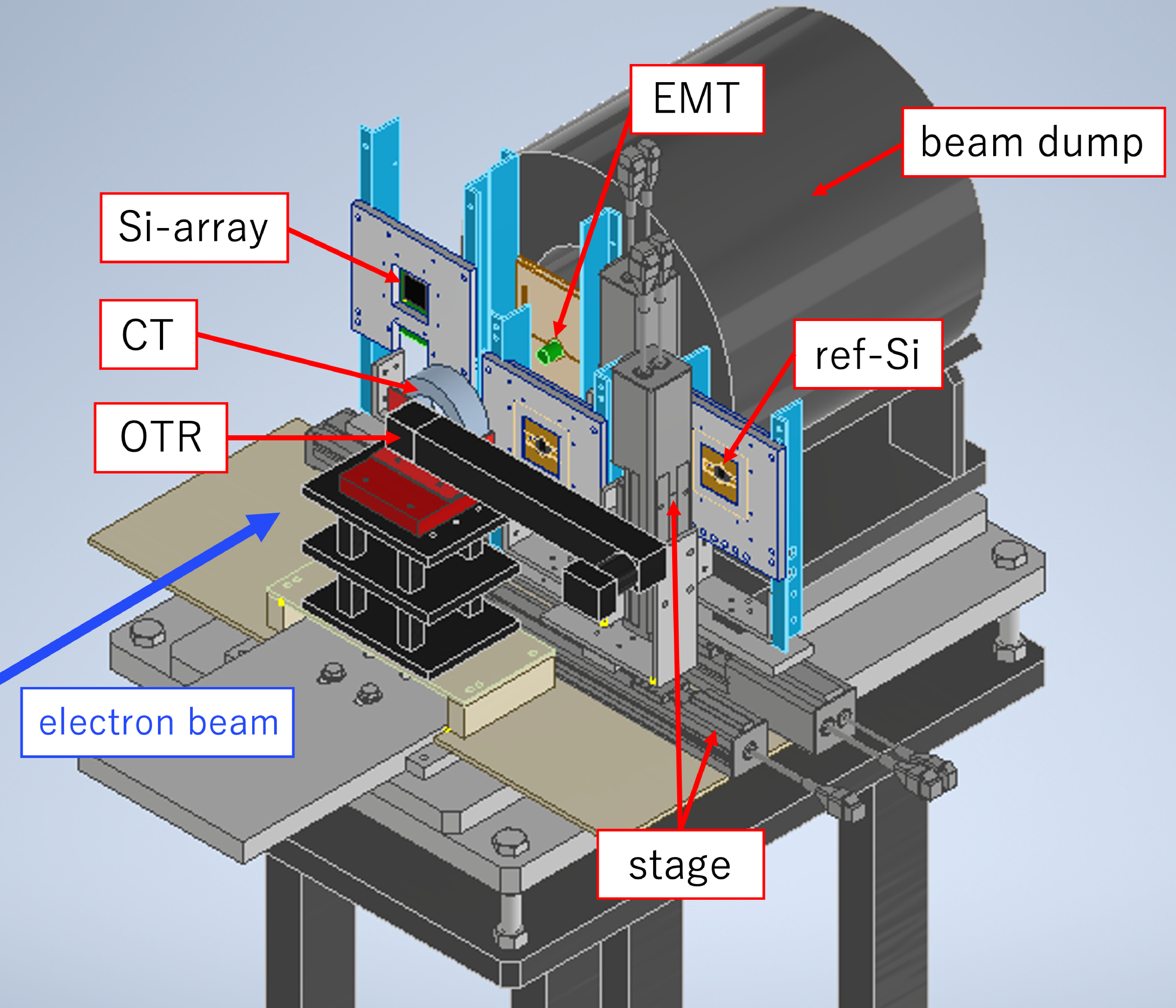}
  \end{minipage}
    \caption{A photograph (left) and conceptual drawing (right) of the experimental setup of the fourth beam test. }
  \label{fig:4thbeam}
\end{figure}

Three different electron beam intensities were adopted depending on the purpose, as shown in Table~\ref{tab:beam}.
We mainly used the low- and high-intensity beams for multiple different purposes.
The low-intensity beam was adjusted to about $1\,\rm pC$ in a macro pulse so that one pulse is equivalent to the muon yield at MUMON for one bunch of the J-PARC proton beam, which is common in the present and future beams. 
The high-intensity beam corresponds to the maximum available intensity of the facility, that is a 20~nC pulse with 7~Hz, and was used to expose EMTs to a large amount of radiation, which is suitable for radiation tolerance investigation.
Another beam setting, the middle intensity, was used only for initial instability investigation.
As with the high intensity beam, the purpose is to expose EMTs to a large amount of radiation, but the amount of electrons in one pulse in the middle-intensity beam is smaller than in the high-intensity beam.
The high-intensity beam operation is aimed at exposing the sensor to charged particles equivalent to an accumulated several hundred days operation of the J-PARC beam line.
The initial instability test using a medium-intensity beam is sufficient for an accumulated 14 day operation of the J-PARC beam line, so a middle-intensity beam was arranged to avoid overexposure.
In order to reduce the effect of electron scattering, energy of the electron beam was set to the highest available one at the facility, 90~MeV.
Several type detectors were used to measure the irradiation amount and the profile for each beam intensity.
Our setup is similar but with some optimization for each beam test, and Figure~\ref{fig:4thbeam} shows the one from the fourth beam test.
In the first beam test, the beam operation was stopped for a few minutes each time the intensity was switched in order to change the setting in the beamline, which might change the vacuum state of the accelerator and cause the beam intensity unstable.
Therefore, in the second beam test, we introduced remote controls for some devices, and in the third and fourth beam test, all settings were controlled remotely.

\begin{figure}[bt]
  \centering
  \includegraphics[width=0.75\linewidth]{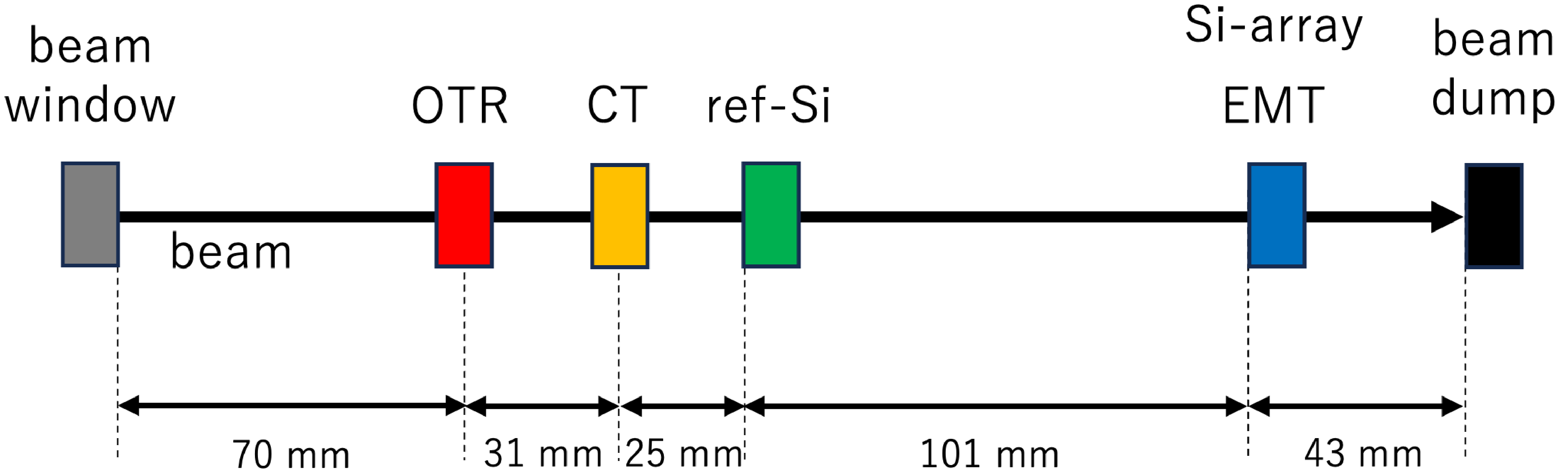}
  \caption{Schematic diagram of the alignment of each detector along the beam axis.}
  \label{fig:detector_pos}
\end{figure}

\begin{table}[bt]
    \caption{Electron beam parameters used in the second to fourth beam tests at RARIS.}
    \label{tab:beam}
    \centering
    \begin{tabular}{lcccc}
        \hline
        \begin{tabular}{c}Beam power\end{tabular}& \begin{tabular}{c}Frequency\\(Hz)\end{tabular}& \begin{tabular}{c}Current \\(nA)\end{tabular}&\begin{tabular}{c}Charge per pulse\\(nC/pulse)\end{tabular} &\begin{tabular}{c}Purpose\end{tabular}\\
        \hline
        Low-intensity&7&0.007&0.001&EMT yield check\\
        Middle-intensity&7& 0.5& 0.071&Irradiation\\
        High-intensity&7&140&20&Irradiation\\
        \hline
    \end{tabular}
\end{table}

\begin{table}[bt]
    \caption{Beam monitor detectors used for the beam irradiation tests at RARIS.}
    \label{tab:detector}
    \centering
    \begin{tabular}{ccll}
        \hline
        Instrument & Model number & Purpose &Beam power\\
        \hline
        ref-Si   & S3590-08      & Intensity monitor & Low  \\
        Si-array & S13620-02     & Profile monitor   & Low    \\
        CT toroidal core & FT-3KM F7555G & Intensity monitor & High / Middle \\
        OTR      & Hand-made     & Profile monitor   & High  \\
        \hline
    \end{tabular}
\end{table}

In order to quantify the amount of iradiation for EMTs, the intensity and profile of the electron beam during the high-intensity operation were measured using a Current Transformer (CT)~\cite{CT_ref} and an Optical Transition Radiation monitor (OTR)~\cite{OTR_T2K}, respectively.
In the low-intensity operation, we measured the beam intensity and profile using a Si pin-photodiode (ref-Si) that is the same type as used in MUMON and a 64-channel silicon pin-photodiode array (Si-array).
The ref-Si is used as a reference to normalize the EMT signal.
The detectors used in the tests are summarized in Table~\ref{tab:detector}.
To minimize the exposure, Si sensor and Si-array were moved to the position away from the beam axis using movable stages during the high-intensity operations.
Figure~\ref{fig:detector_pos} schematically shows the locations of the OTR, CT, ref-Si, Si-array, and EMT.
In the fourth beam test, a heater and a thermometer were installed to measure the temperature dependence of the EMT response.
The CT, EMT, and Si signals were acquired with an 8-channel 250-MHz sampling 14-bit 2-Vpp Flash Analog to Digital Converter (FADC) (CAEN DT5725).
The Si-array signal was acquired using two 32-channel 62.5-MHz sampling 12-bit 2-Vpp FADCs (CAEN DT5740).

\subsection{High- and middle-intensity beams}\label{sec:subsection_high_and_middle_intensity}
\begin{figure}[bt]
  \centering
\includegraphics[width=0.6\linewidth]{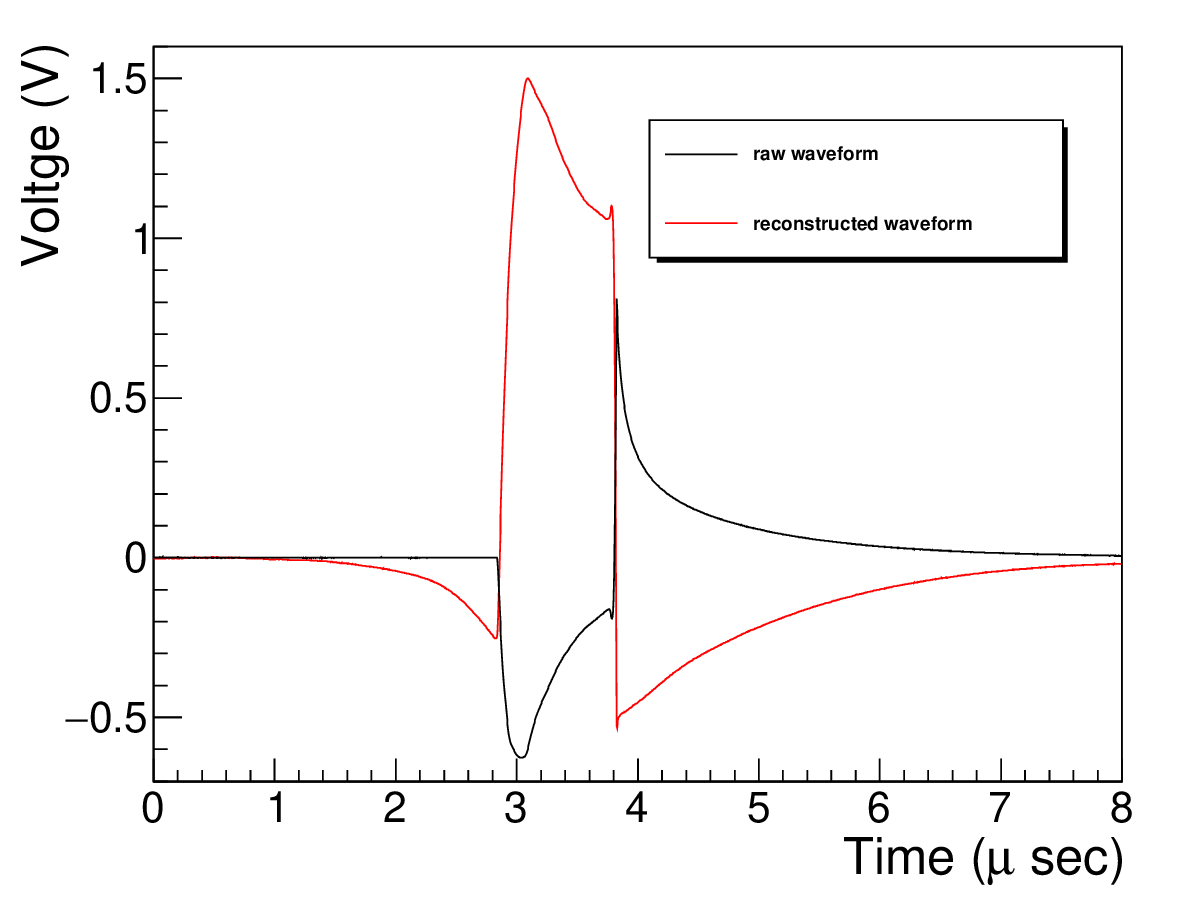}
  \caption{Example of the raw CT waveform and reconstructed longitudinal beam shape at a high-intensity beam setting. Charge in this pulse is 32.97 nC.\label{fig:CT_waveform}}
\end{figure}

During the high-intensity beam operation, the EMT signal line and ground were connected by a 1$\,\rm\Omega$ resistor to suppress the voltage drop by too much current.
The CT has a ring core whose inner and outer diameters are $79\,\rm mm$ and $51\,\rm mm$, respectively, and width is $25\,\rm mm$.
The material is FINEMET (Hitachi Metals, Ltd.) with a magnetic permeability of $18500\,\rm H/m$. 
The number of turns of the coil is 1 which makes the CT response differential.
This was decided to get a sufficiently large pulse height.
The frequency-dependent response was measured in advance and was used to reconstruct the beam longitudinal distribution from the CT output waveform.
The beam flux was calculated by integrating the area of the reconstructed waveform.
Typical raw and reconstructed waveforms are shown in Figure~\ref{fig:CT_waveform}.
As can be seen around the 2$\sim$3 $\rm\mu$s and 4$\sim$7 $\rm\mu$s of the reconstructed waveform, the baseline distortion remained and reconstruction is not as good as desired.
This is because of the low sampling rate compared to the beam pulse.
The flux of each beam pulse was calculated from the product of the height and width of the reconstructed waveform.

\par
\begin{figure}[bt]
    \centering
    \includegraphics[width=0.6\textwidth]{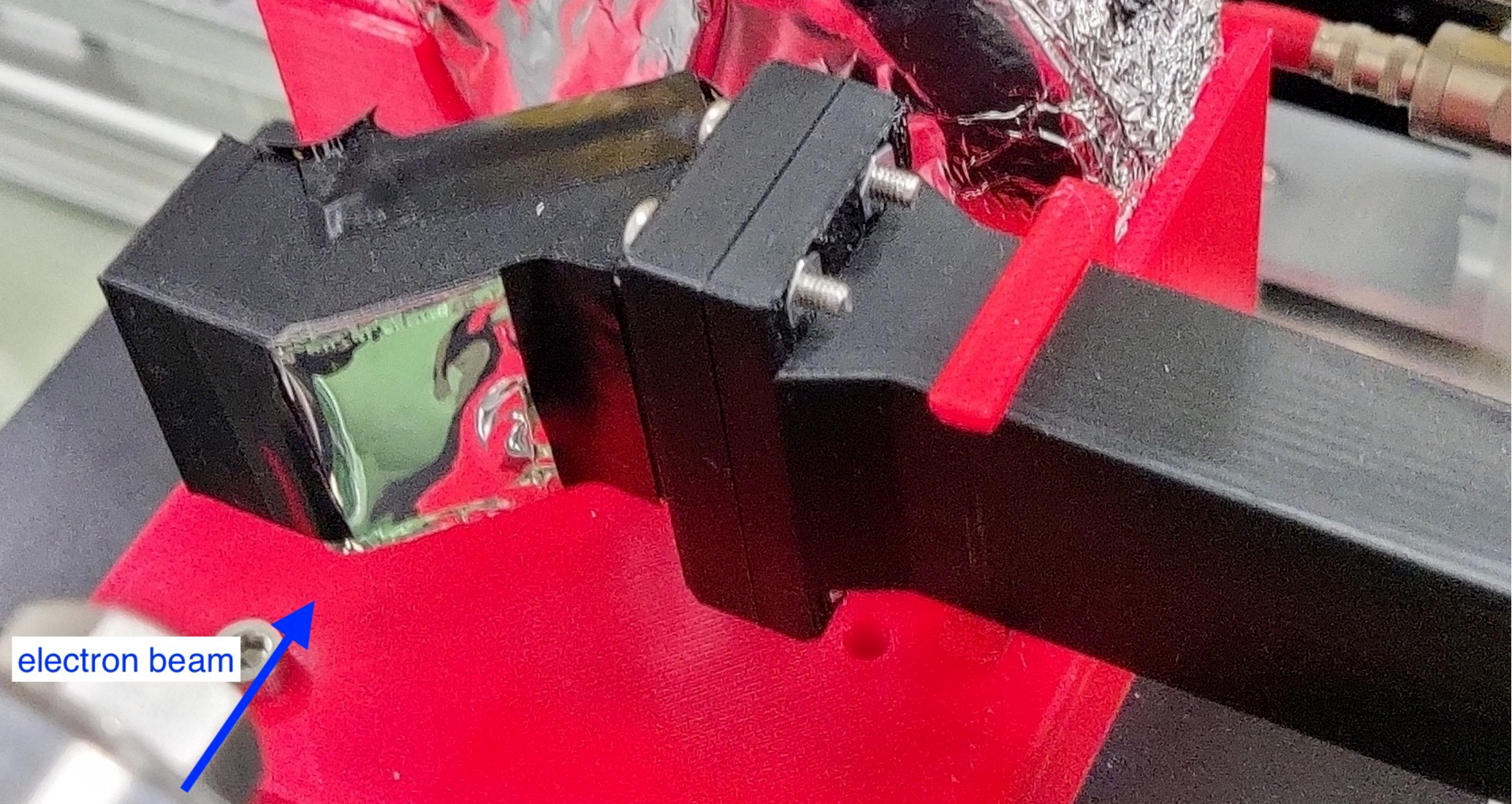}
    \caption{Photograph of the OTR foil. The aluminium foil is attached to a 1 cm square window.}
    \label{fig:otr_foil}
\end{figure}
\begin{figure}[bt]
    \centering
    \includegraphics[width=0.5\textwidth]{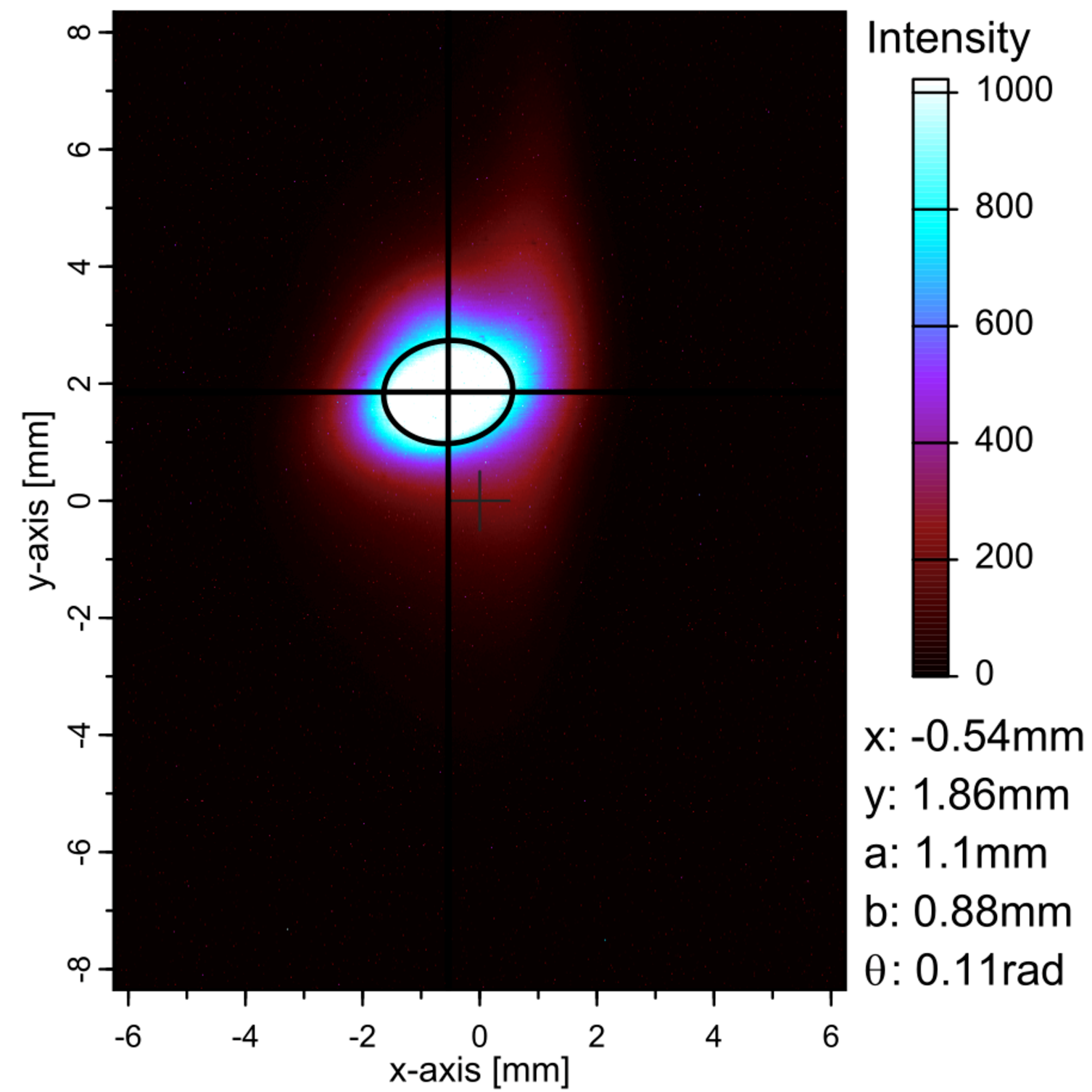}
    \caption{High-intensity electron beam profile for one bunch taken with OTR.
    The beam profile analysis was accomplished by determining the beam center ($x$, $y$ in the legend) and finding an ellipse with its two axes ($a$, $b$) describing the respective 1$\rm \sigma$ distribution of the fit. $\rm \theta$ indicates the angle between the $x$ axis of the OTR and the $a$ axis of the ellipse, for $\theta=0$~rad it holds that $a = 2 \sigma_{x}$ and $b = 2\sigma_{y}$.}
    \label{fig:otr_pro}
\end{figure}
\begin{figure}[bt]
    \centering
    \includegraphics[width=0.7\textwidth]{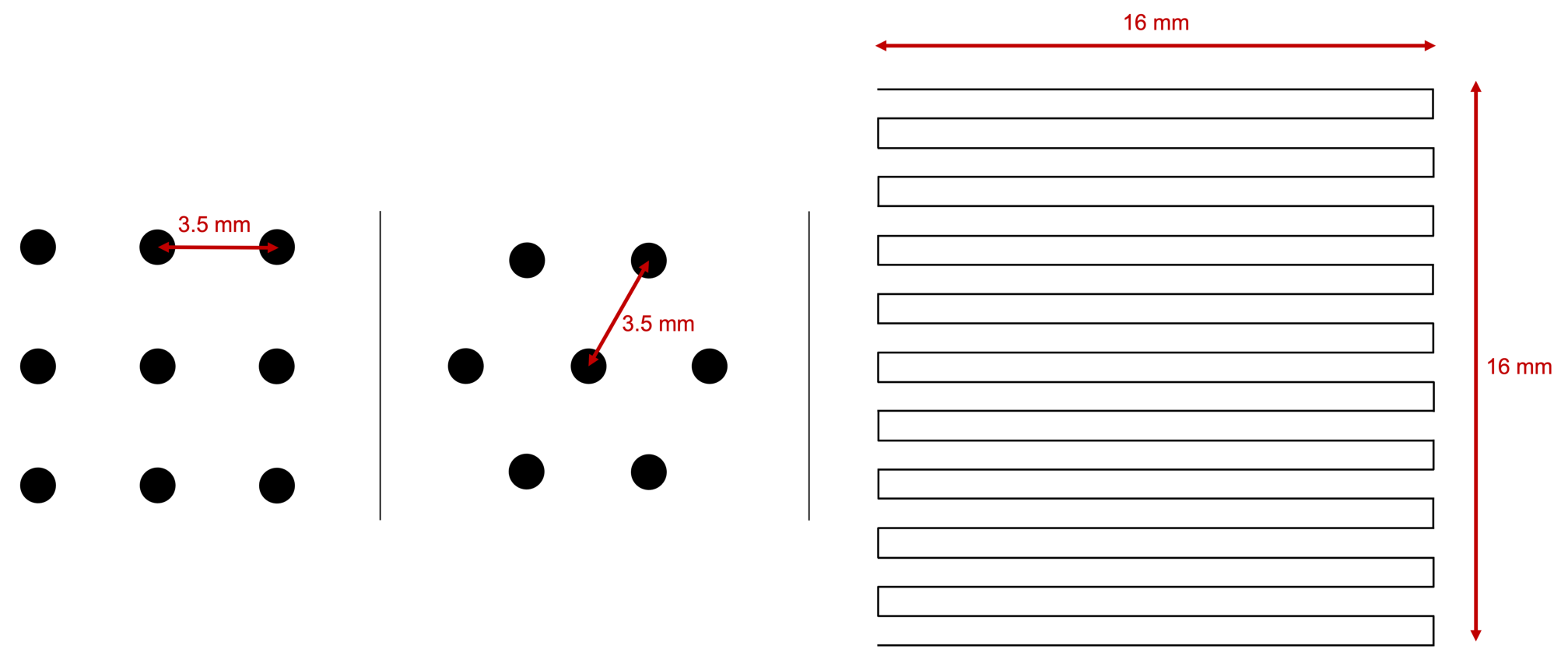}
    \caption{Multiple points irradiation in a high-intensity beam. 
    The black circles in the left panel represent the irradiation points in the second beam test and the middle panel is for those in the third beam test.
    The right panel shows the moving lines in the fourth beam test.}
    \label{fig:beam_pos_high-intensity}
\end{figure}

The OTR monitor is used since the third beam test as a specialized profile monitor for the high-intensity beam.  
An aluminum foil was set at 45 degrees to the beam axis as the OTR generation medium (see Figure~\ref{fig:otr_foil}), and images were taken with a CMOS camera (model number CS165MU, lens: FL-BC7528-9M) positioned 30$\,\rm cm$ away.
Figure~\ref{fig:otr_pro} shows a typical beam profile taken by OTR during the high-intensity beam operation.
The irradiation efficiency is defined as the ratio of electrons that hit the EMT out of the electrons in the beam.
The electron beam size of RARIS is about 1\ mm which is smaller than the sensitive area of EMTs.
The muon beam size of the J-PARC neutrino beamline is a few hundred cm and then expected to irradiate EMTs almost uniformly.
We designed an experiment to achieve a uniform irradiation on EMTs by changing their position with a movable stage during the high-intensity beam operation.
In the second beam test, EMTs were irradiated at nine points in a grid pattern, and in the third beam test, EMTs were irradiated at seven points in a hexagonal arrangement, as schematically shown in Figure~\ref{fig:beam_pos_high-intensity}.
From the third beam test, we installed OTR to measure the high-intensity beam profile, and calculated the irradiation efficiency which is defined as the fraction with which beam particles hit the EMT's sensitive area.
The irradiation efficiency at the third beam test is found to be 70\%.
However, it is also found that the irradiation density differed by a factor of 3 depending on the position on the EMT.
In the fourth beam test, in order to improve the uniformity, each EMT was irradiated while continuously scanned over a range of $16\times 16\,\rm mm^2$ as shown in the right-hand side of Figure~\ref{fig:beam_pos_high-intensity}.
The resultant irradiation efficiency is about 28\%, and the non-uniformity is about 3\% or less.
By multiplying the irradiation efficiency to the beam flux measured by CT, the total exposure on each EMT is determined.

\subsection{Low-intensity beam}\label{sec:subsection_low_intensity}
\begin{figure}[bt]
  \begin{minipage}[b]{0.45\linewidth}
    \centering
    \includegraphics[width=1.0\linewidth]{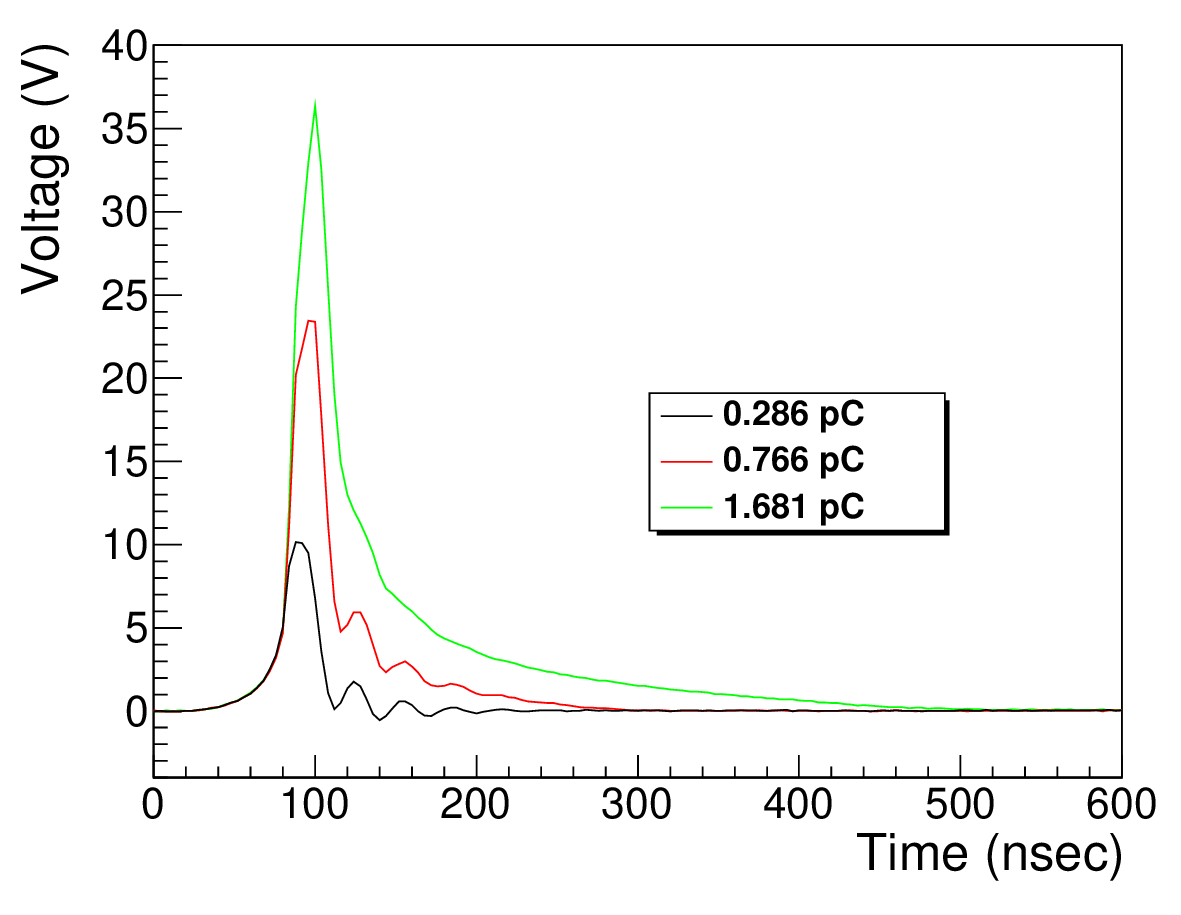}
  \end{minipage}
  \begin{minipage}[b]{0.45\linewidth}
    \centering
    \includegraphics[width=1.0\linewidth]{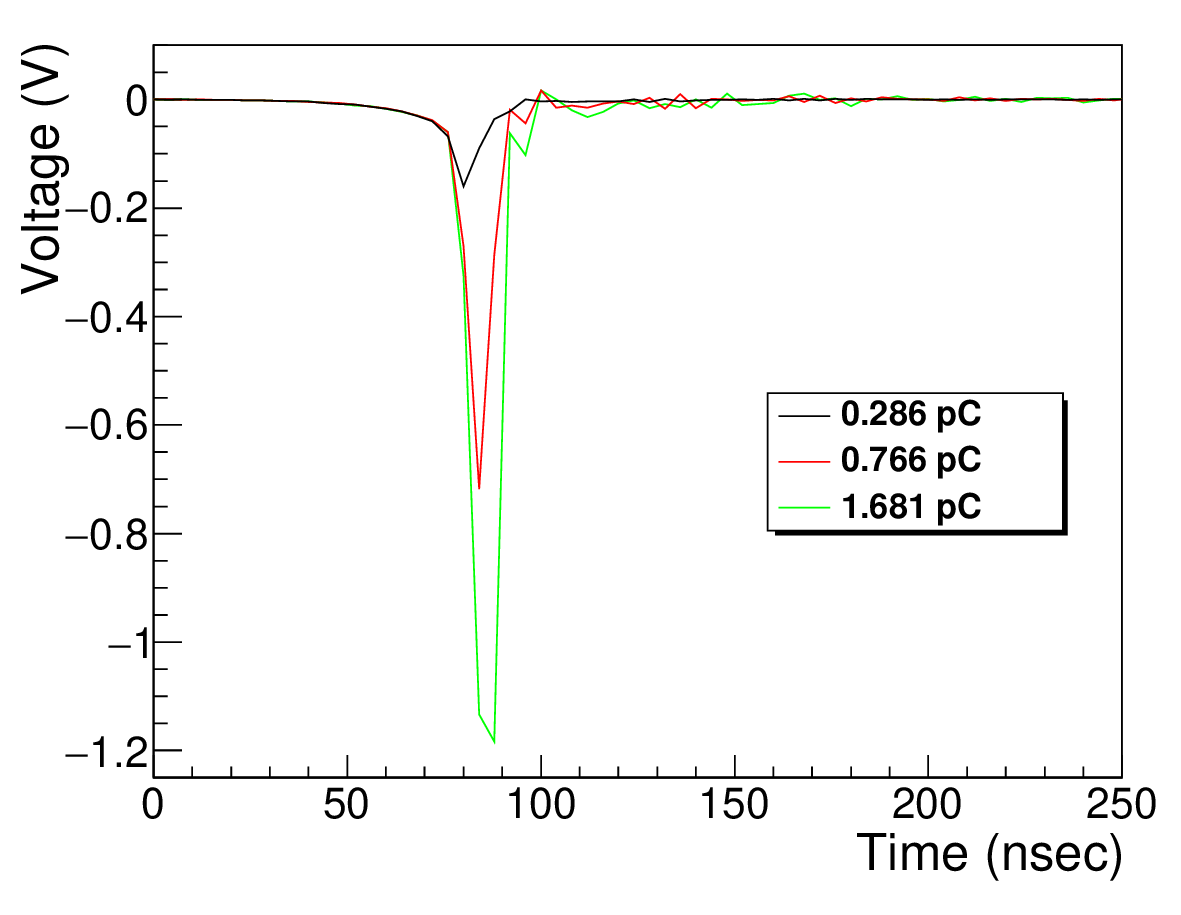}
  \end{minipage}
  \caption{Example waveforms from the ref-Si (left) and EMT (right) in the low-intensity beam operation.\label{fig:si_emt_waveform}}
\end{figure}

The ref-Si was used to measure the intensity of the low-intensity beam.
An attenuator with a factor of 1/79.4 was applied.
A low-pass filter was inserted in the bins line to ensure stable voltage supply for Si sensors.
When measuring the EMT yield, we acquired the ref-Si signal at the same time, and by taking the ratio, we can accurately measure the EMT yield regardless of shot-to-shot variations in the electron beam.
The applied voltage to EMTs was $-450\,\rm V$ for the signal check.
Typical ref-Si and EMT waveforms are shown in Figure~\ref{fig:si_emt_waveform}.

\begin{figure}[bt]
    \centering
    \includegraphics[width=0.75\linewidth]{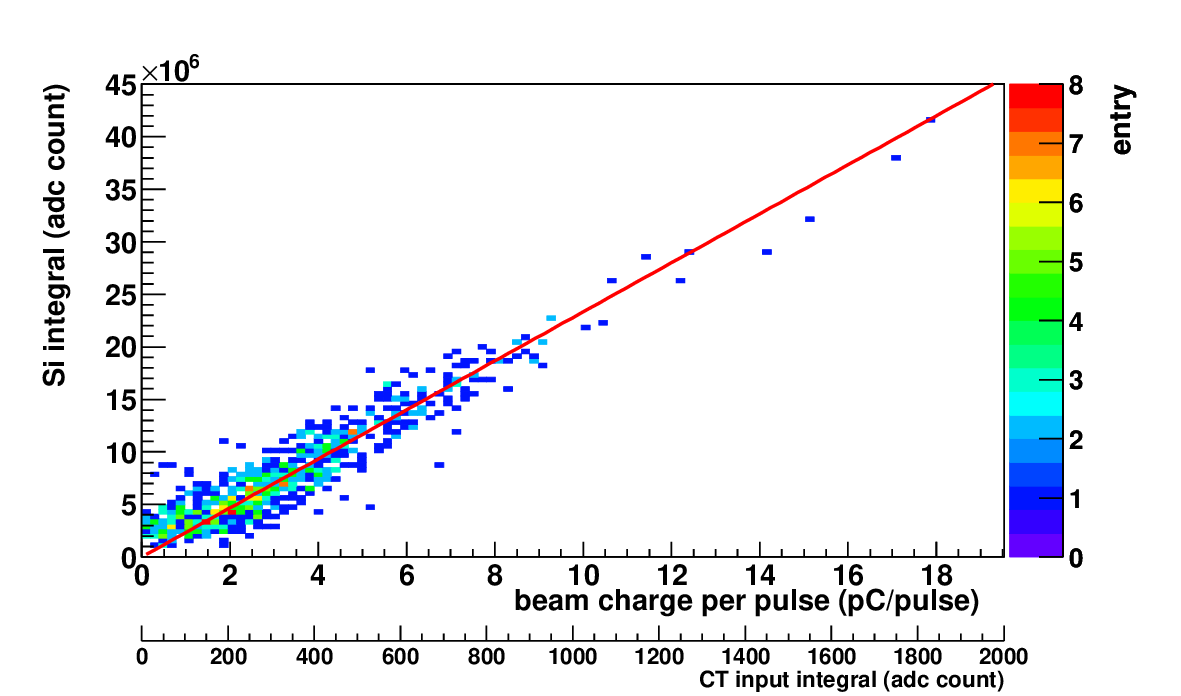}
    \caption{Relation of integrated charge signals between ref-Si and CT. The slope of the linear function fit is $2.3\times 10^9\,\rm ADC\, counts\cdot pluse/nC$.}
    \label{fig:ct_vs_refsi}
\end{figure}

The ref-Si was calibrated using the CT.
Since the low-intensity beam is too weak to be detected by the CT, the calibration was performed using a beam that is several times more intense than the low-intensity beam.
Figure~\ref{fig:ct_vs_refsi} shows the relation between the ref-Si signal yield and beam intensity measured by the CT.


\par
\begin{figure}[bt]
  \centering
  \includegraphics[width=0.9\linewidth]{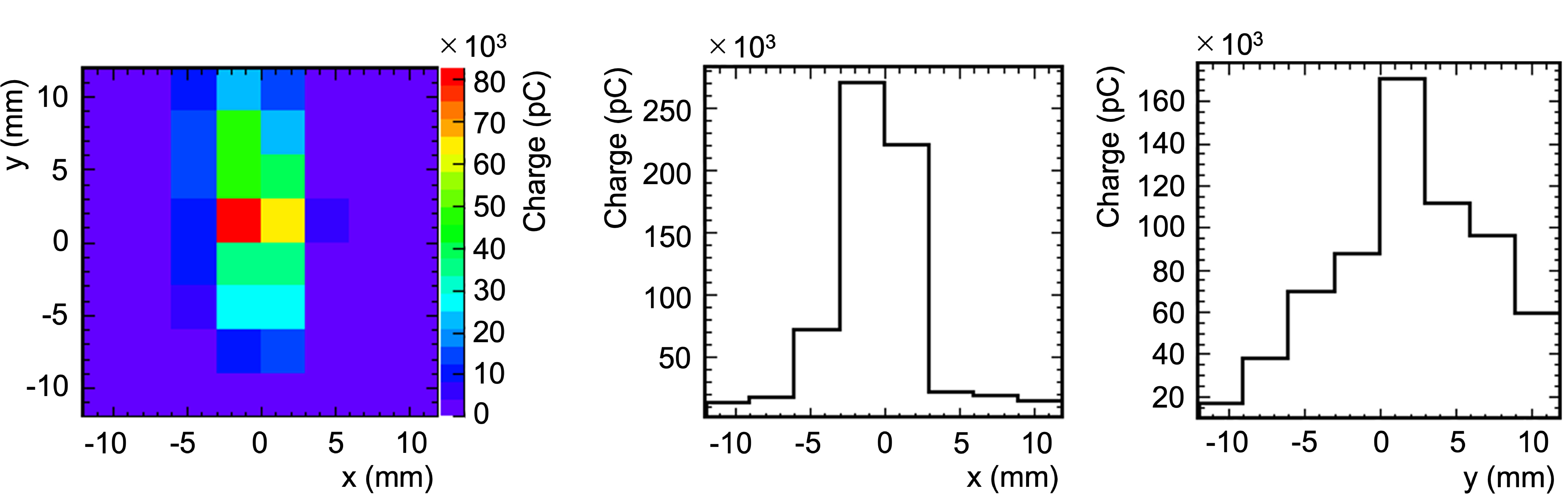}
  \caption{Typical low-intensity beam profile measured by the Si-array in the fourth beam test. 
  The measured beam position with its width (mean and standard deviation) is ($-0.44\pm 3.51\,\rm mm, 1.85\pm 5.26\,\rm mm$).
  }
  \label{fig:siarray}
\end{figure}

The Si-array was used to measure the profile of the low-intensity beam.
The Si-array has 64 $(= 8\times 8)$ silicon pin-photodiode sensors whose size is $\rm2.5\, mm$ at $\rm3\, mm$ intervals and can monitor a $23.5 \times 23.5\,\rm mm^2$ area.
The variation in gain for each channel was measured in advance using an LED.
We calculated the signal charge for each channel of the Si-array and reconstructed a two-dimensional profile as shown in Figure~\ref{fig:siarray}.
By fitting this with a two-dimensional Gaussian, the position and size of the beam are obtained.
The fluctuation in the position of the low-intensity beam is approximately $1\,\rm mm$ in horizontal direction and $4\,\rm mm$ in vertical direction, and a typical beam width is 1$\sim$6~mm which is smaller than the EMT cathode area.

\section{Results}\label{sec:section_Result}
\subsection{Linearity}\label{sec:subsection_linearity}
\begin{figure}[bt]
  \begin{minipage}[b]{0.45\linewidth}
    \centering
    \includegraphics[width=1.15\linewidth]{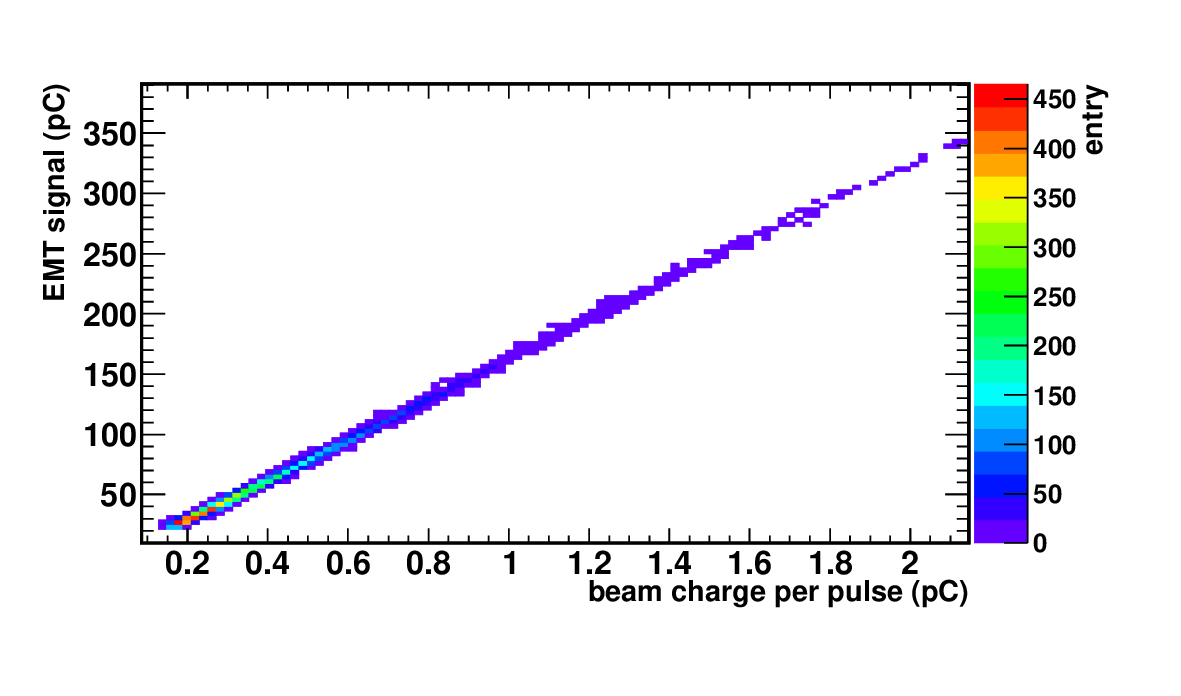}
  \end{minipage}
  \hspace{+10truept}
  \begin{minipage}[b]{0.45\linewidth}
    \centering
    \includegraphics[width=1.15\linewidth]{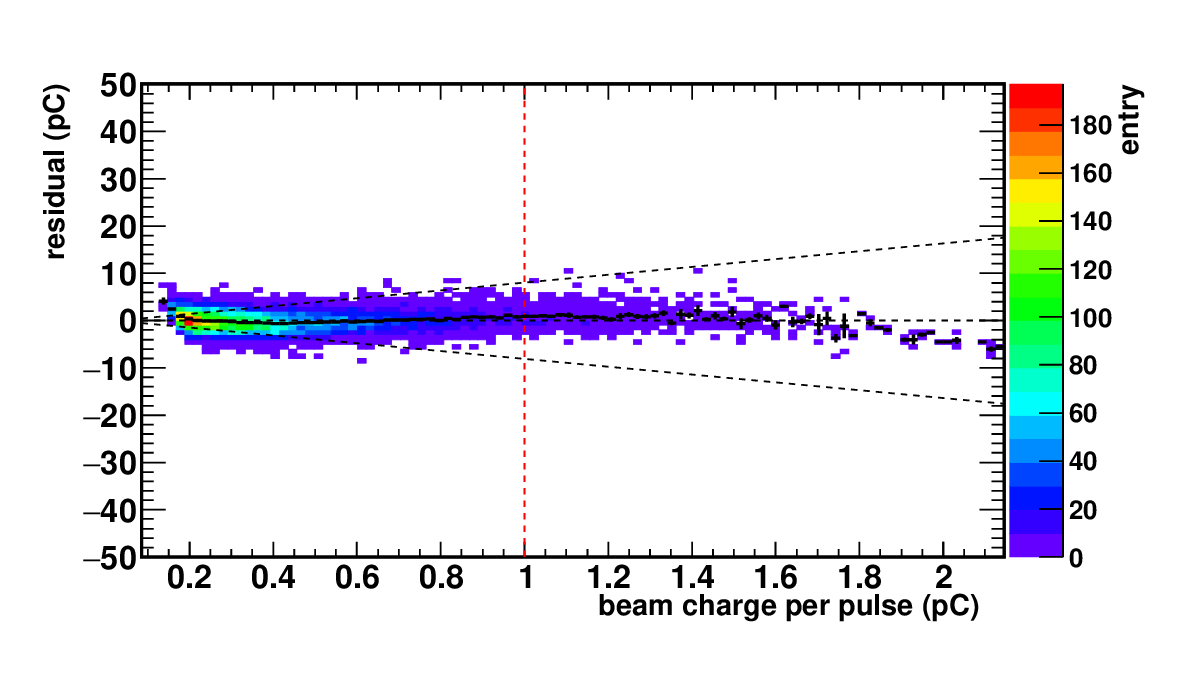}
  \end{minipage}
  \caption{Measured integrated charge signals from EMT as a function of the incident beam power (left). 
  The slope from a linear function fit is 166.1$\pm$0.5. The figure in the right panel shows the residuals from the fitted straight line. Here, the black dotted lines indicate the zero and $\pm$5\% of the value in the vertical axis. The average of the y-axis values for each bin on the x-axis is shown as a black line. The vertical red dotted line shows the charge of one bunch exposed to EMTs in the J-PARC beamline in 2028.\label{Fig::EMT_linearity}}
\end{figure}

In order to evaluate the linearity of EMTs, we irradiated the similar flux as from the J-PARC neutrino beamline operation and compared the yield with Si sensors (see Figure~\ref{Fig::EMT_linearity}).
According to the previous studies~\cite{MUMON2010}, Si sensors keep their linearity within $\pm 1.9\%$ up to $3.5\,\rm pC/cm^2$.
For the linearity measurement, instead of setting certain beam powers that requires special care in beam control and hence is difficult, we made use of pulse-to-pulse variations of the low-intensity beam. 
The expected muon charge per pulse from the J-PARC neutrino beam at the 1.3~MW operation is 1.0 pC for each EMT.
The beam irradiation test result demonstrates that the EMT response is linear within $\pm$5\% for the range between 0.2 and 1.0$\,\rm pC/pulse$, the region of interest in the present study.






\subsection{Radiation tolerance}\label{sec:subsection_radiation_tolerance}
\begin{figure}[bt]
\centering
\includegraphics[width=12cm]{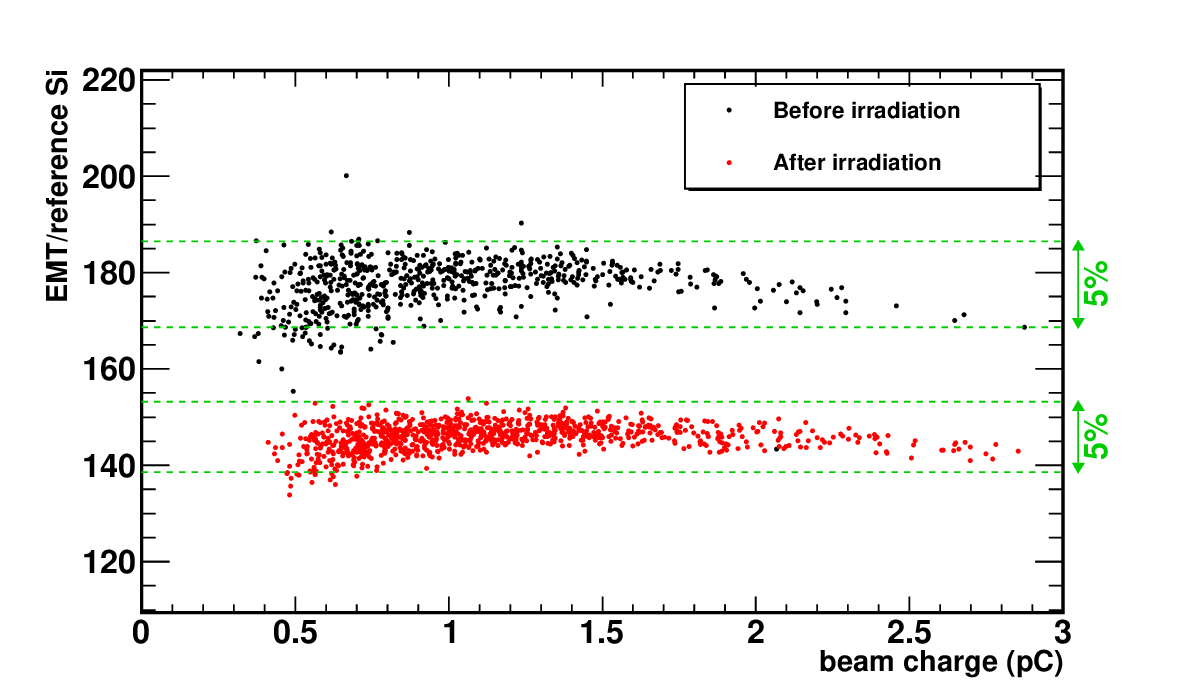}
\caption{\label{fig:linearity_multi}The EMT integrated charge signal before (black points) and after (red points) irradiation with a $\rm 450\, nC/cm^{2}$ electron beam. This measurement was performed with varying charges around 1 pC, which covers the region of interest in the present study. 
The green dotted lines represent the 5\% ranges for each.
}
\end{figure}

\begin{figure}[bt]
\centering
\includegraphics[width=14cm]{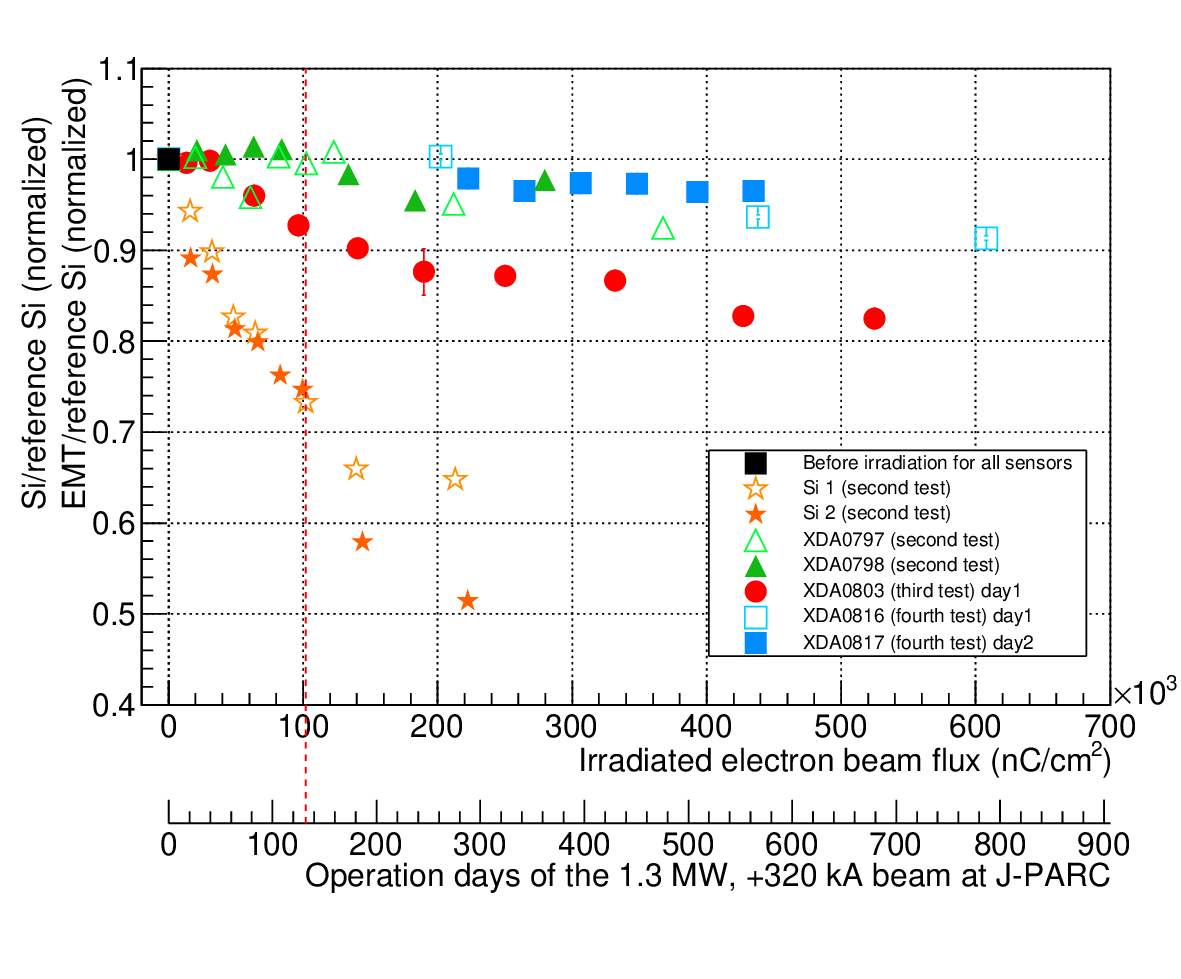}
\caption{\label{fig:radiation_tolerance}
The relative yield of the EMT and Si detectors to ref-Si normalized to the initial state. 
The horizontal axis corresponds to the flux of the electron beam irradiated to the sensors and the value converted into the number of days at the J-PARC neutrino beam operation for the MUMON center.
The flux is assumed for the sensitive area, which is $\rm 0.77\, cm^2$ for the EMT and $\rm1\, cm$ for the Si, and for the MUMON center.
The vertical axis is the relative integrated charge signal amount when the yield before radiation irradiation is set to 1. 
The black point represents a reference point for all sensors.
The ref-Si was not irradiated with a high-intensity beam.
Different colors are used to indicate different beam tests as well as different tested sensors.
The vertical red dotted line corresponds to 132 days of the future operation.}
\end{figure}

In order to confirm whether the EMT can be used for a long period at the J-PARC neutrino beamline, we irradiated EMTs with a high-intensity electron beam equivalent to several years of the J-PARC neutrino beam operation, and measured the relative integrated charge yield of EMTs.
As shown in Figure~\ref{fig:linearity_multi}, the EMT integrated charge signal is decreased, while linearity keeps within 5\%
after electron irradiation at $\sim$$\rm 450\, nC/cm^{2}$ that corresponds to a $\sim$580 days operation at J-PARC in 2028.
Figure~\ref{fig:radiation_tolerance} shows the irradiation results obtained from the second to fourth beam tests. 
For comparison, the Si sensor results are also shown. 
The radiation tolerance of the EMT is overwhelmingly better than that of the Si, and even after 700 days of operation at the future power of 1.3 MW, the integrated charge signal decrease is less than 18\%.
It was found that there are individual differences in EMTs. Some EMTs only degrade by 5\% after 550 days of irradiation at 1.3~MW.

\par
The typical operating period for the beam in a year anticipated at Hyper-Kamiokande is 132 days, and therefore we require that the sensors withstand at least 132 days of a continuous operation.
If the Si is operated for 132 days at the future beam condition, the integrated charge yield would decrease by 25\%, and in addition, the leakage current would increase, making it impossible to apply high voltage.
On the other hand, the integrated charge signal decrease of the EMT is expected to be less than 8\% even after 132 days.
Even some EMTs hardly degraded after 132 days.
The degradation rate was found to be slower after more than 200 days of use.
Therefore, by calibrating regularly, it can be used for several years without replacement.

Systematic errors on the result in Figure~\ref{fig:radiation_tolerance} are estimated and shown in Table~\ref{table:systematic_signal_check}.
The uncertainty on the EMT yield normalized by ref-Si arises from the variations in beam position because the sensitive areas of the EMT and Si are different. 
The beam profiles measured by Si-array were used to estimate fluctuation of the beam entering the sensitive areas.
We estimated the systematic uncertainty of the CT charge calculation method. 
This is the uncertainty that arises when converting the irradiation amount to the number of operation days in the J-PARC neutrino beamline.
The MUMON sensors are exposed to muons with an average energy around $\rm3\, GeV$, while the beam is composed of $90\,\rm MeV$ electrons in this beam test.
Simulations with FLUKA~\cite{FLUKA1, FLUKA2} show that the energy deposit to an EMT is different by about 3\%.
Therefore, when converting the exposure to the EMT in this beam test to that of the J-PARC neutrino beam operation, this 3\% is taken as a systematic uncertainty.
For the low-intensity beam operation, the effect of particle difference is cancelled out because the evaluation is based on the EMT yield over the Si yield.

\begin{table}[bt]
  \caption{Systematic uncertainties on the measured flux of the irradiated beam. Since the measurements were performed for two days in the third and fourth beam tests, systematic uncertainties were evaluated for each day. The result of the first beam test is not shown here as it has not been quantitatively evaluated.}
  \label{table:systematic_signal_check}
  \centering
  \begin{tabular}{lccc}
    \hline
     & \multicolumn{3}{c}{Systematic uncertainty (\%)}  \\
    \multicolumn{1}{l}{Uncertainty source}  & Second & Third (Day1, Day2) & Fourth (Day1, Day2) \\
    \hline 
    Beam fluctuations & $\pm$0.6 & $\pm$4.1, $\pm$3.9 & $\pm$2.1,  $\pm$4.6\\
    Reproduction of CT input signal&$^{+2.1}_{-10}$&$\pm$13&$\pm$13\\
    Particle difference & $\pm$3 & $\pm$3 & $\pm$3 \\
    \hline
  \end{tabular}\end{table}





\subsection{Signal stability}\label{sec:signal_stability}
\begin{figure}[bt]
\centering
\includegraphics[width=0.75\textwidth]{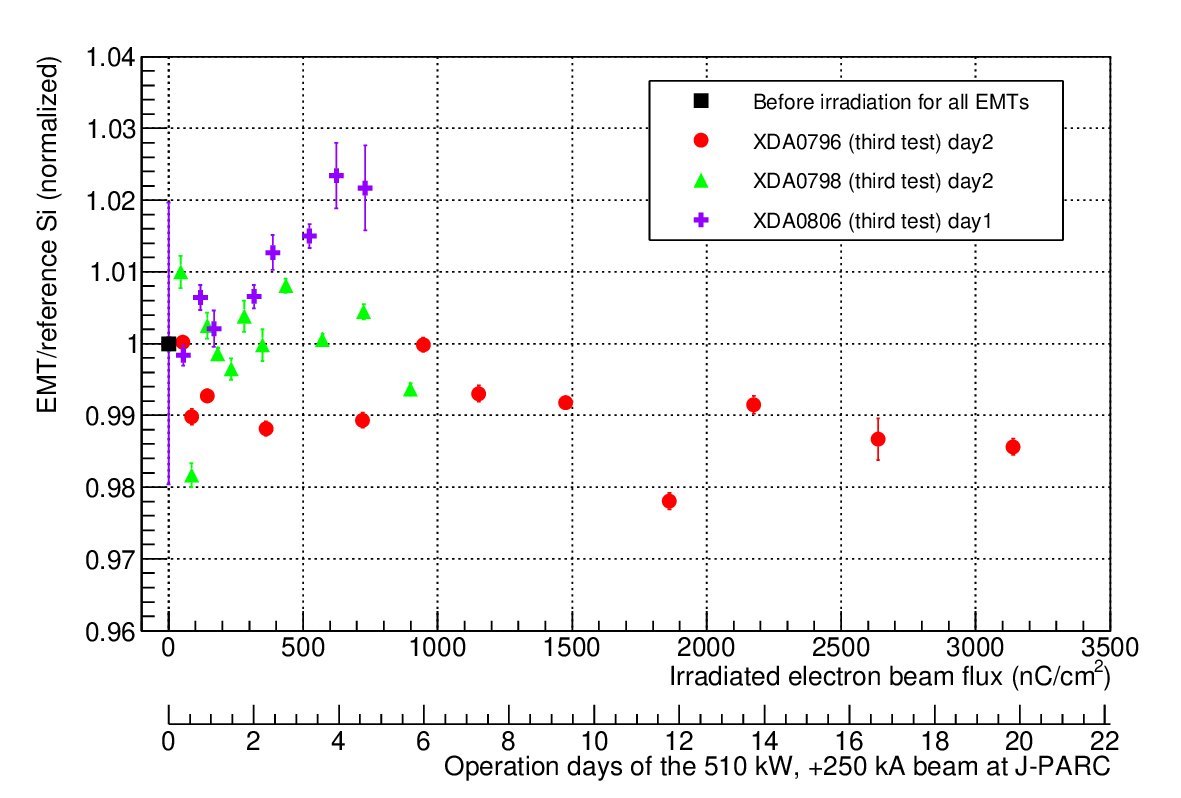}
\caption{\label{fig:initial_instability_kakkokari}
Variation in the EMT integrated charge signal as a function of exposure.
The ref-Si was not irradiated with a middle-intensity beam.
The vertical axis shows the relative ratio with the first reference point that is a black point for all EMTs as 1.
The legend notes the serial number of each EMT and the date of the beam test. 
The systematic errors for this result is same as those described at the end of the Section~\ref{sec:subsection_radiation_tolerance}.}
\end{figure}

In the measurement at the MUMON site in the past, the integrated charge yield decrease by 2.6\% was observed about a week after the start of irradiation, and then the yield became stabilized, as reported in Ref.~\cite{Ashida2018}.
In order to confirm the stability of EMTs at the beginning of the operation in T2K, we repeated the process of an exposure equivalent to several days of T2K operation using the middle-intensity setting (see Table~\ref{tab:beam}) and measured the EMT response before irradiation with the high-intensity beam. 
The measurement results are shown in Figure~\ref{fig:initial_instability_kakkokari}.
The result indicates that EMTs have individual differences in performance, which is part of our study items in the beam test.
In this measurement, the variation in the integrated charge yield during the middle-intensity irradiation is less than 2\%.
This will be further discussed in Section~\ref{sec:subsection_temperature}.



\section{Discussion}\label{sec:section_discussion}
In this section, we will discuss the following two points.
One is the temperature dependence as a potential cause for the initial instability observed at J-PARC.
The other is the cause of the integrated charge signal decrease of up to 18\% by the irradiation as shown by the red points in Figure~\ref{fig:radiation_tolerance}.

\subsection{Temperature dependence}\label{sec:subsection_temperature}
As described in Section~\ref{sec:signal_stability}, initial instability was not observed as clearly as the J-PARC test~\cite{Ashida2018}.
During the test at J-PARC, EMTs were installed at location where the temperature is not controlled and can vary as much as $\rm15\, K$ on a weekly time scale depending on the accelerator operation status.
Therefore, we measured the temperature dependence of the EMT integrated charge signal in the fourth beam test.
Unlike the test location at J-PARC in Ref.~\cite{Ashida2018}, the actual MUMON sensor location is surrounded by the enclosure and temperature is controlled at approximately $34^\circ\rm C$, so the temperature dependence measurements were performed including $34^\circ\rm C$.
\par
\begin{figure}[bt]
    \centering
     \includegraphics[width=10cm]{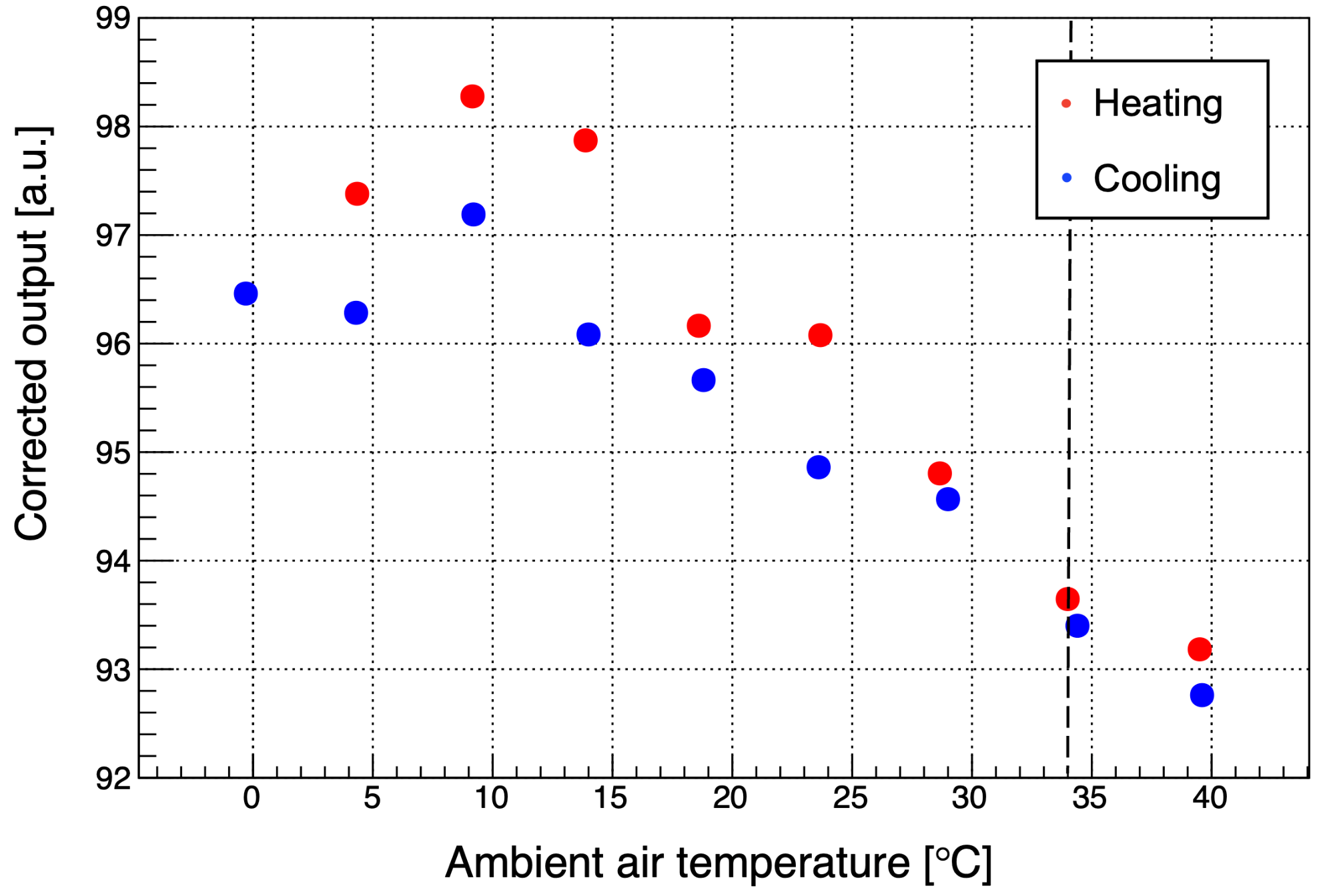}\caption{Measured temperature dependence of the EMT integrated charge signal yield using an LED exposure. 
    The vertical axis shows the value of the EMT output corrected for the LED temperature dependence.
    Red points are data when the temperature is rising, and the blue points are data when the temperature is falling. The EMT integrated charge signal is corrected by the temperature dependence of the LED emission.
    The dotted line corresponds to the temperature that is controlled at the MUMON sensor location ($34^\circ\rm C$).
    }\label{fig:EMT_temp_LED}
\end{figure}

Although the EMT cathode is made of vapor-deposited aluminum, it is sensitive to photons to some extent, so we evaluated using an LED at first.
The EMT was placed in a constant temperature bath and the yield was measured at different temperatures.
Since the LED response also changes depending on the temperature, we corrected the LED response based on the standard specification ``NSPW500GS-K1'' of Nichia Chemical Industries, Ltd. as a typical temperature characteristics.
At the same time, we also measured the Si, which is supposed to have small temperature dependence of the signal\footnote{The temperature dependence of the Si was found to be $\pm0.2\%$ in the range $0^\circ\rm C$ to $40^\circ\rm C$.}, to confirm the correction of the temperature characteristics of the LED.
Figure~\ref{fig:EMT_temp_LED} shows the integrated charge yield at different temperatures. 
The temperature dependence coefficient of the EMT yield is 0.16--0.18\%/$^\circ \rm C$, which is insufficient to explain the integrated charge signal fluctuations at around the MUMON test site.

\par
\begin{figure}[bt]
\centering
\includegraphics[width=10cm]{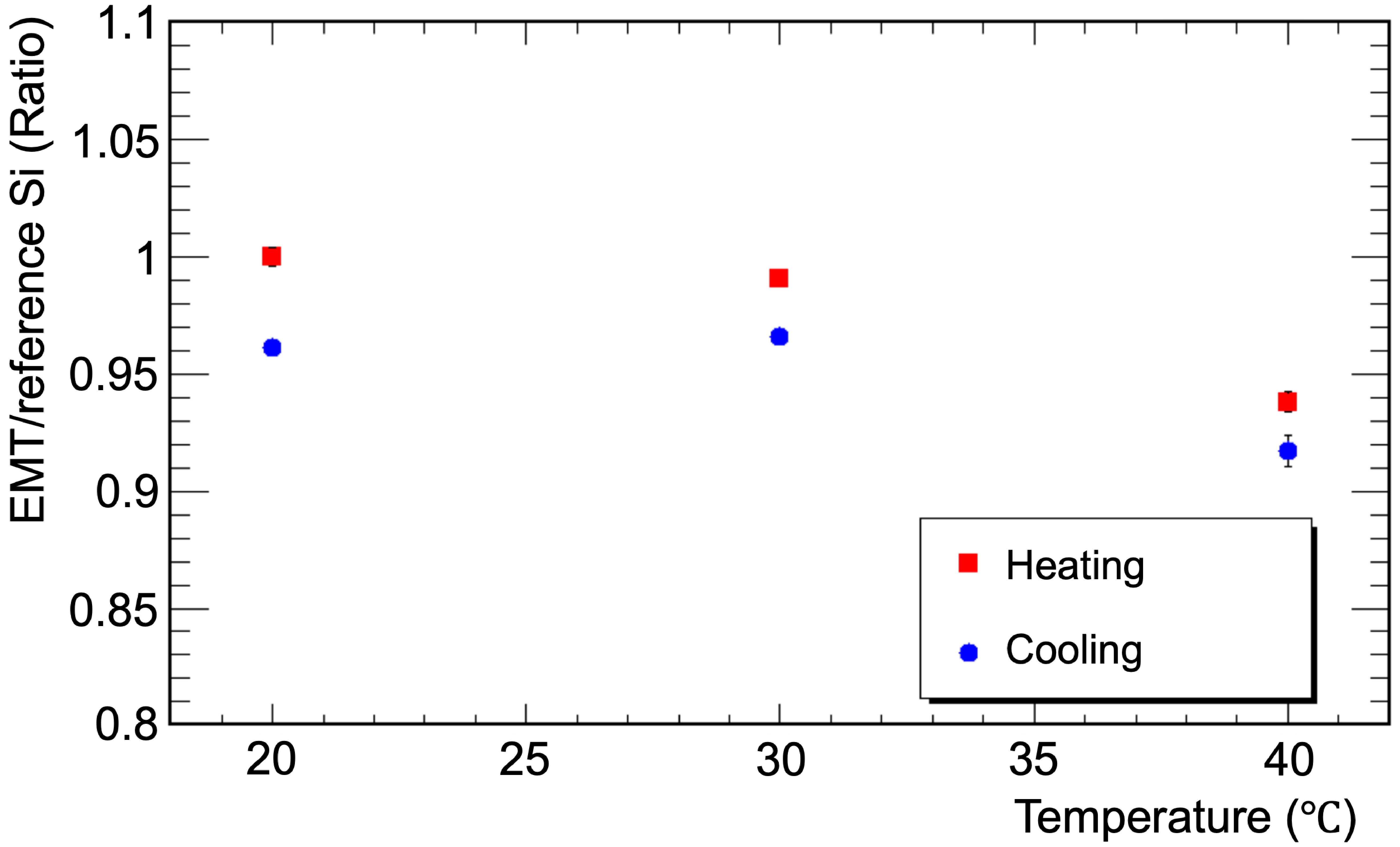}
\caption{\label{fig:temp_dep_beam} Temperature dependence of  the EMT integrated charge yield measured at fourth beam test. The measurement started at $20\rm ^\circ C$, increased to $40\rm ^\circ C$ (red points), and then decreased to $20\rm ^\circ C$ (blue points). The vertical axis is the ratio to the yield at the first $20\rm ^\circ C$.}
\end{figure}

We measured the temperature dependence of the EMT integrated charge signal also in the fourth beam test.
The temperature was varied by applying hot air to the EMT using a heater and fan, and a platinum resistance thermometer was used to measure the temperature at the EMT location.
Figure~\ref{fig:temp_dep_beam} shows the measured temperature dependence, and the temperature change coefficient in the range of 30$^\circ \rm C$ or higher is $0.6\%/^\circ\rm C$.
This is larger than that measured using an LED.
\par
Among the measurements using the LED and that of the electron beam irradiation test, the latter measures the signal generated penetrating charged particles. 
The reason of different temperature dependence is unknown.
We consider the latter case to estimate the effect at J-PARC because the signal generation mechanism is same.
Then, the yield instability of 2--3\% corresponds to a variation of about $4^\circ\rm C$ and can explain the observed initial instability at J-PARC.
The temperature inside the temperature-controlled enclosure, which is the actual environment of the MUMON sensor location, is controlled so that the temperature change is less than $\pm 0.7^\circ\rm C$ \cite{MatsuokaDtheses}. Therefore, the temperature dependence of the EMT is not an issue even if it exists.


\subsection{Cause for yield decrease}\label{sec:subsection_cause_of_radiation_degradation}
\begin{figure}[bt]
\centering
\includegraphics[width=9cm]{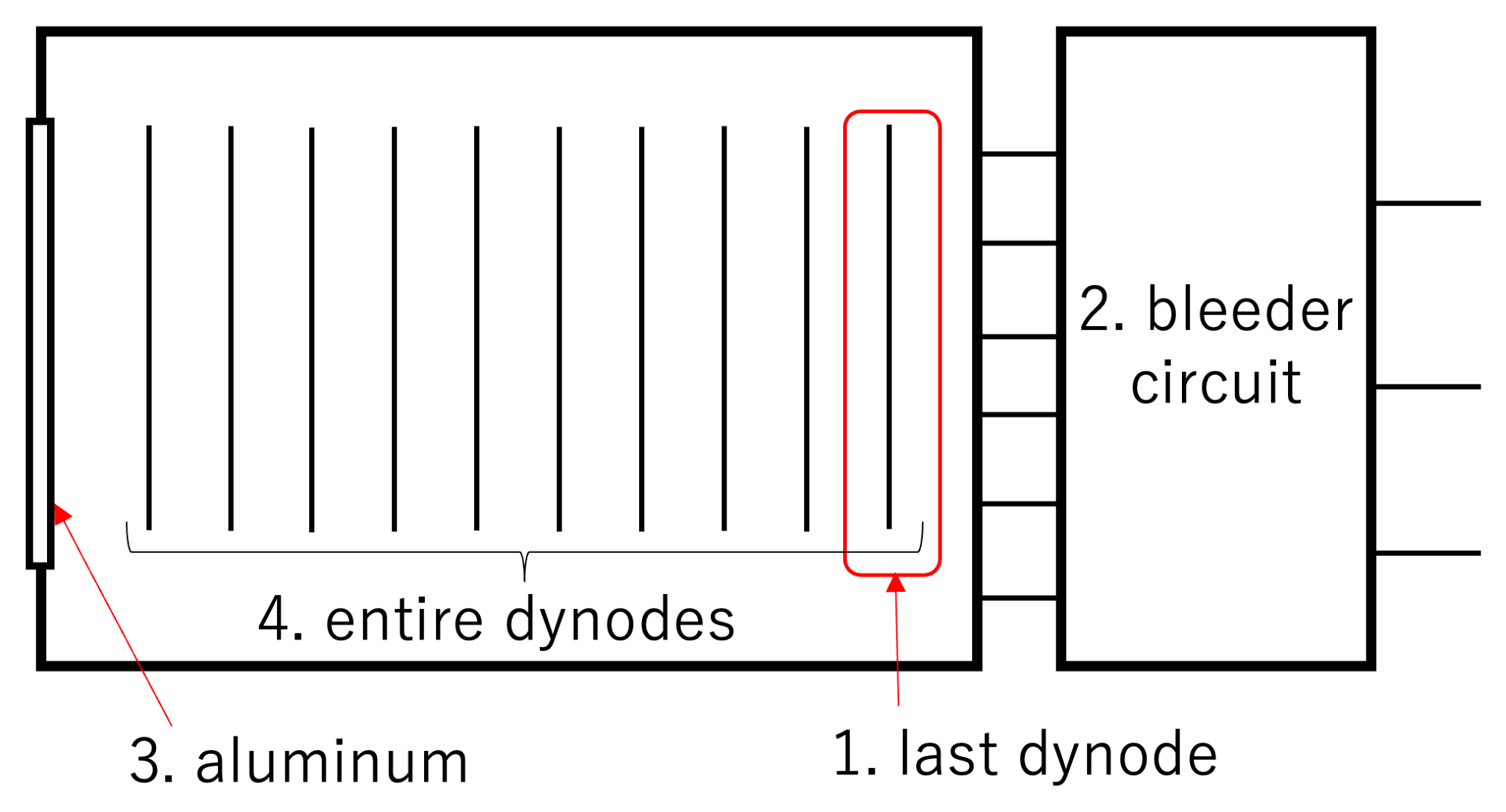}
\caption{\label{fig:cause_of_signal_degradation} Schematic diagram of an EMT. Numbers are assigned to each location where the effects of radiation damage are discussed in Sections~\ref{sec:subsubsection_Last_dynode} through \ref{sec:subsubsection_Entire_dynodes}.}
\end{figure}

The radiation tolerance of EMTs is found to be much better than that of the presently used Si, but an integrated charge signal drop up to $18\%$ was observed when irradiated with $700\times10^3\,\rm nC$ electrons.
If the cause of the degradation is identified, EMTs could be used for longer by taking countermeasures.
In order to find out where the degradation originates, we investigate each of the four components of the EMT: the last dynode, bleeder circuit, aluminum cathode, and entire dynodes.
Each location is schematically shown in Figure~\ref{fig:cause_of_signal_degradation}.


\subsubsection{Last dynode}\label{sec:subsubsection_Last_dynode}
\begin{figure}[bt]
\centering
\includegraphics[width=12cm]{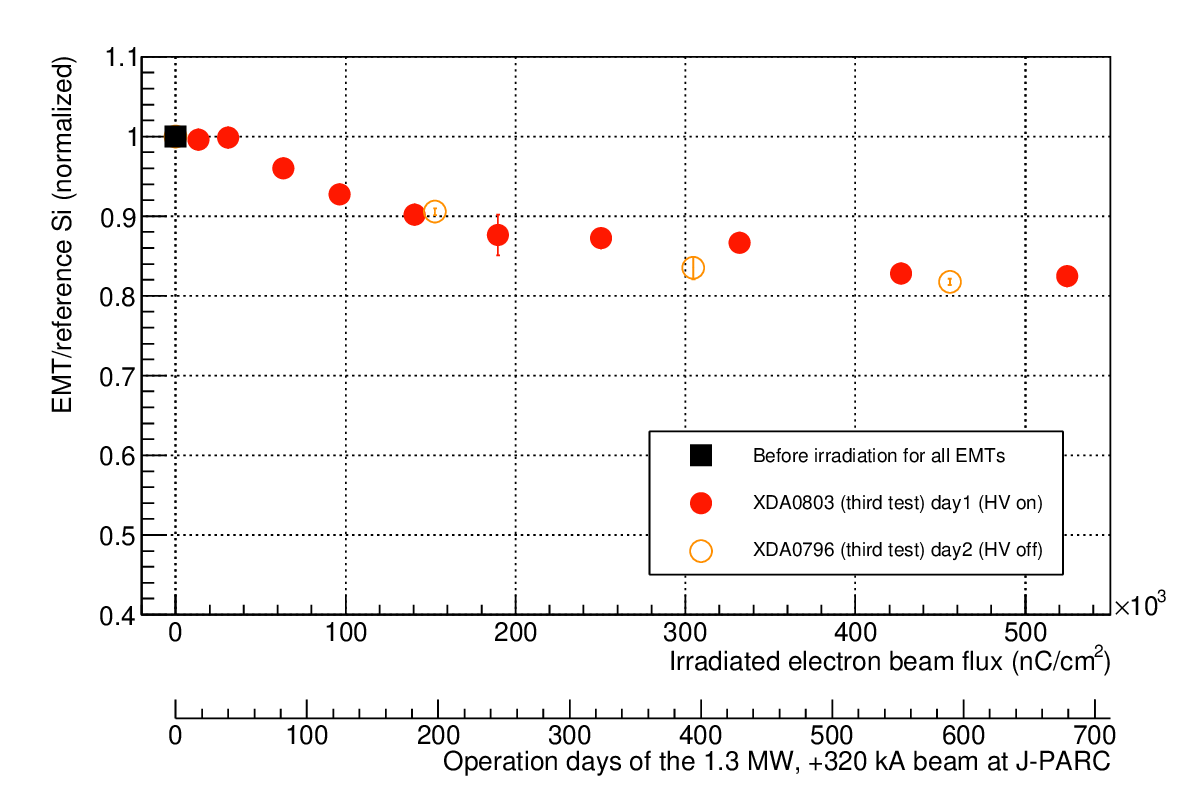}
\caption{\label{fig:3rd_HV_onoff} EMT yield as a function of radiation dose with and without high voltage applied measured in the third beam test. The axes are same as Figure~\ref{fig:radiation_tolerance}. The legend notes the serial number of each EMT, identification and date of the test.
The systematic errors for this result is same as those described at the end of the Section~\ref{sec:subsection_radiation_tolerance}.}
\end{figure}

With amplification of electrons, a large amount of secondary electrons collide with the final stage of the dynode.
In order to verify the hypothesis that the integrated charge signal is reduced due to the last dynode degradation, we conducted an exposure test with and without high voltage application in the third beam test.
Figure~\ref{fig:3rd_HV_onoff} shows the result from this test.
Even when high voltage was not applied during exposure, the similar degradation was observed.
This result indicates that degradation in the later stage of the dynodes is not a main cause of radiation damage.

\subsubsection{Bleeder circuit}\label{sec:subsubsection_Bleeder_circuit}
We irradiated only the bleeder circuit with a high-intensity beam and then attached to an EMT to measure the change in yield. 
The result is summarized in Table~\ref{tab:bleeder_check}.
There observed a tendency for the integrated charge yield yield to slightly increase rather than decrease.
We measured the resistance and capacitance of the circuits before and after irradiation using a multimeter.
There is no significant difference in the measured values and they are consistent with the catalog values.


\begin{table}[bt]
  \caption{The EMT yield change after irradiating only the bleeder circuit with a high-intensity beam of $2.5\times10^5$--$4.7\times10^5\,\rm nC$ (equivalent to J-PARC $\rm1.3\, MW$ 330--620 day). The ratios of the yield before and after the irradiation are shown.}
  \label{tab:bleeder_check}
  \centering
  \begin{tabular}{lcc}
    \hline
    Beam test & \begin{tabular}{c}Irradiation amount\\(nC)  \end{tabular}& \begin{tabular}{c}Change of the yield\\(normalized)\end{tabular} \\
    \hline
    Third test (Day 3)  & $2.5\times 10^5$ & 1.1   \\
    Fourth test (Day 2) & $4.7\times 10^5$ & 1.01    \\
    Fourth test (Day 3) & $3.3\times 10^5$ & 1.025  \\
    \hline
  \end{tabular}
\end{table}

\subsubsection{Aluminum cathode}\label{sec:subsubsection_Aluminum_cathode}
We investigated the possibility that the probability of electron emission from the aluminum-deposited cathode surface was worsened by irradiation.
We prepared a special bleeder circuit in which the resistance between the cathode and the first dynode ($R_1$ in Figure~\ref{fig:div_cir}) is set to $0\,\Omega$.
When this bleeder circuit is used, when a charged particle passes through, the initial electron is not generated at the cathode but at the first dynode.
Therefore, if the cathode degradation is the main cause of the integrated charge signal decrease, the special circuit would prevent the integrated charge yield from decreasing.
Table~\ref{tab:cathode_check} shows the yield change after the high-intensity beam irradiation for each condition.
There are variations in the result, but the yield reduction is observed for one of the tests with the special bleeder circuit.


\begin{table}[bt]
  \caption{Survey results regarding degradation of aluminum cathode. Yield ratio shows the relative yield after irradiating only the EMT body with a high-intensity beam with respect to the yield before irradiation.}
  \label{tab:cathode_check}
  \centering
  \begin{tabular}{lccc}
    \hline
    Beam test & \begin{tabular}{c}Irradiation amount\\(nC) \end{tabular}& Bleeder circuit & \begin{tabular}{c}Change of the yield \\(normalized) \end{tabular}\\
    \hline
    Fourth test (Day 2) & $4.7\times 10^5$ & normal   &  0.92\\
    Fourth test (Day 2) & $4.7\times 10^5$ & special  &  0.91\\
    Fourth test (Day 3) & $3.3\times 10^5$ & normal   & 1.045\\
    Fourth test (Day 3) & $3.3\times 10^5$ & special  & 1.001\\
    \hline
  \end{tabular}
\end{table}


\subsubsection{Entire dynodes}\label{sec:subsubsection_Entire_dynodes}
Since the three hypothesis discussed above are found to be not likely the major cause of degradation, we suspect that the degradation of dynodes would be the main cause.
The state of the secondary electron emitting agent (alkali-metal antimony) coated on the dynodes might have been deteriorated due to irradiation, and the secondary electron amplification factor has decreased.
In Figure~\ref{fig:radiation_tolerance}, EMTs that have decreased by only a few~\% can be explained by the Aluminum cathode, while it is not enough to explain the EMT that has decreased by 18\%, hence a remaining possibility exists with the entire dynodes.
We prepared EMTs without the secondary electron emitting agent and conducted a test. 
However, the gain of such EMTs was found to be very low even when applying a voltage of $\rm-1000\, V$: about 0.5\% compared to a usual EMT with an applied voltage of $\rm-450\, V$.
The radiation tolerance of that EMT was not measured, as no signal was seen in the low-intensity setting of the beam irradiation test.
The signal is too small to accurately measure the muon beam and we need an alternative approach to suppress the integrated charge signal decrease if it is due to degradation of the dynodes.




 
\section{Summary}\label{sec:section_summary}
We have been developing EMTs as a new sensor for the muon monitor in the future J-PARC neutrino beamline operation.
We measured the linearity and radiation tolerance of EMTs using the electron beam at RARIS.
The EMTs show linearity better than $\pm 5\%$ up to the future beam intensity.
The integrated charge yield decreases by 8\% over the equivalent 132 days of the 1.3 MW beam operation, indicating that EMTs have much higher radiation tolerance than the currently used Si sensors.
After over 200 days of use with the 1.3~MW operation at J-PARC, the degradation becomes rather mild, so with careful calibration, it may be possible to keep using EMTs without replacement.
In the beam test, the cause of the degradation was investigated in detail.
The aluminium cathode, bleeder circuits and last dynodes are found to be not likely the main cause.
The most suspicious cause is the dynode degradation.
We also investigated the temperature dependence of the EMT response. 
The measured temperature dependence of the EMT integrated charge yield indicates the initial instability previously observed in the prototype test at J-PARC is due to the variation of temperature at the test location, while this should not be an issue as the actual muon monitor location will be temperature controlled.
This study confirms that EMTs retain a good linearity response and high radiation tolerance compared to the present Si sensors. 
From the reported results, we concluded that EMTs can be used as a new MUMON detector and are currently preparing for the actual installation in the J-PARC neutrino beamline. 

\section*{Acknowledgments}
We thank Hamamatsu Photonics K. K. for the cooperation in producing the EMTs.
This study was performed using facilities of Research Center for Accelerator and Radioisotope Science, Tohoku University (Proposal No. 2915, 2943, 2977, 3007).
The authors are grateful to the J-PARC accelerator group for supplying a stable beam.
This work was partially supported by JSPS KAKENHI Grant Numbers JP17J06141, JP16H06288, JP21K03591 and the U.S.-Japan Science and Technology Cooperation Program in High Energy Physics.
This work was supported by JST, the establishment of university fellowships towards the creation of science technology innovation, Grant Number JPMJFS2138.




\end{document}